\documentclass{emulateapj}
\usepackage{amssymb,amsmath,latexsym,natbib,wasysym}
\usepackage{pdfcolmk}

\DeclareMathAlphabet{\mathpzc}{OT1}{pzc}{m}{it}

\newcommand\be{\begin{equation}}
\newcommand\ee{\end{equation}}

\def\x{\times}
\def\e#1{10^{#1}}

\begin{document}

\pagenumbering{arabic}

\slugcomment{Accepted by the Astrophysical Journal}
\shorttitle{Probing Extra-solar Planet Interiors}
\shortauthors{Ragozzine \& Wolf}

\title{Probing the Interiors of Very Hot Jupiters Using
  Transit Light Curves}

\author{Darin Ragozzine \& Aaron S. Wolf\footnote{Both authors
contributed equally to this work.}}
\affil{Division of Geological and Planetary Sciences, California Institute
of Technology, Pasadena, CA 91125}

\email{darin@gps.caltech.edu, awolf@gps.caltech.edu}

\begin{abstract} 
Accurately understanding the interior structure of extra-solar planets 
is critical for inferring their formation and evolution. The internal density
distribution of a planet has a direct effect on the star-planet
orbit through the gravitational quadrupole field created by the rotational and tidal
bulges. These quadrupoles induce apsidal precession that is proportional to 
the planetary Love number ($k_{2p}$, twice the
apsidal motion constant), a bulk physical characteristic of the planet that 
depends on the internal density distribution, including the presence or absence 
of a massive solid core. We find that the quadrupole of the planetary
tidal bulge is the dominant source of apsidal precession for very
hot Jupiters ($a \lesssim 0.025$ AU), exceeding the effects of general relativity
and the stellar quadrupole by more than an order of magnitude. 
For the shortest-period planets, the planetary interior
induces precession of a few degrees per year. 
By investigating the full photometric signal of apsidal precession, we find that changes
in transit shapes are much more important than transit timing variations. 
With its long baseline of ultra-precise photometry,
the space-based \emph{Kepler} mission can
realistically detect apsidal precession with the accuracy necessary to 
infer the presence or absence of a massive core in very hot
Jupiters with orbital eccentricities as low as $e \simeq 0.003$. The signal 
due to $k_{2p}$ creates unique transit light curve variations that are generally not 
degenerate with other parameters or phenomena.
We discuss the plausibility of measuring $k_{2p}$ in an effort to 
directly constrain the interior properties of extra-solar planets. 
\end{abstract}


\maketitle

\section{Introduction}
Whether studying planets within our solar system or planets orbiting other stars, 
understanding planetary interiors represents our best
strategy for determining their bulk composition, internal dynamics,
and formation histories. For our closest neighbors, we have had the luxury of sending spacecraft to
accurately measure the higher-order gravity fields of these objects,
yielding invaluable constraints on their interior density distributions. Using
these observations, we have been able, for instance, to infer the presence of
large cores, providing support for the core-accretion theory of planet 
formation \citep{2005AREPS..33..493G}. Study of planets outside our solar system, however, has
necessitated the development and usage of more indirect
techniques. Nevertheless, as the number of well-characterized
extra-solar planets grows, we gain more clues that help us answer the
most fundamental questions about how planets form and evolve.

Guided by our current understanding of planetary physics, we have begun
to study the interiors 
of extra-solar planets. This endeavor has been dominated by a
model-based approach, in which the mass and radius of a planet are
measured using radial velocity and transit photometry observations,
and the interior properties are inferred by finding the model most
consistent with those two observations. This strategy clearly requires
a set of assumptions, not the least of which is that the
physical processes at work in extra-solar planets are just like those
that we understand for our own giant planets. While it does seem that
this approach is adequate for explaining most of the known transiting
planets, there does exist a group of planets for which the usual set of
assumptions are not capable of reproducing the observations 
\citep[e.g.,][]{2006A&A...453L..21G,2007ApJ...661..502B}.
These are the planets with so-called
positive ``radius
anomalies'',  including the first-discovered transiting planet
HD 209458b \citep{2000ApJ...529L..45C}.  Though most of these planets can be explained by
adjusting different pieces of the interior physics in the models (including
opacities, equations of state, and heat deposition), it is currently impossible to 
discern which combination of these possible explanations is actually
responsible for their observed sizes \citep{2006A&A...453L..21G}.

Additional uncertainties also exist for planets at the other end of
the size spectrum. For the group of under-sized extra-solar planets, such as HD 149026b,
 the canonical approach is to give the planet a
massive highly condensed core of heavy elements in order to match the
observed radius. This approach also provides a first order estimate of
the planet's bulk composition, in terms of its fraction of heavy
elements. There is
also the added complication of how the assumed state of
differentiation affects the inferred composition and predicted
structure \citep{Baraffe2008}. 

Currently, the most promising approach to modeling the distinctive features of extra-solar planet interiors
is to study the known transiting planets as an ensemble. The group can be used to develop 
either a single consistent model that reproduces all the observations \citep[e.g.,][]{2006A&A...453L..21G} 
or to showcase the possible diversity in model parameters \citep[e.g., opacities, as in][]{2007ApJ...661..502B}. 
Surely, a model-independent measure of interior structure would be valuable in order to begin disentangling 
otherwise unconstrained physics. 

The idea of obtaining direct structural measurements for distant objects is
by no means a new one. For decades, the interiors of eclipsing binary
stars have been measured by observing ``apsidal motion,''
i.e. precession of the orbit due to the non-point-mass component of
the gravitational field \citep{1928MNRAS..88..641R,1938MNRAS..98..734C,1939MNRAS..99..451S,1939MNRAS..99..662S}. 
The signal of the changing orbit is encoded in
the light curves of these systems by altering the timing of the
primary and secondary eclipses. From these eclipse times, it is
straightforward to determine the so-called apsidal motion 
constant which then constrains the allowed interior density distributions.
Interior measurements inferred
from apsidal precession were among the first indications that stars
were highly centrally condensed. While it seems non-intuitive, we show
in this paper that we can use a similar technique to measure the
interior properties of very hot Jupiters. Most surprisingly, the
interior structure signal for very hot Jupiters actually 
dominates over the signal from the star, yielding an unambiguous
determination of planetary interior properties.

Our theoretical analysis is also extended to full simulated photometry in order to
explore the observability of apsidal precession. 
We show that this precession is observable
by measuring the subtle variations in transit light curves. The photometric analysis is focused on the data expected from 
NASA's \emph{Kepler} mission, which successfully launched on March 6, 2009 
\citep{2003SPIE.4854..129B,2006Ap&SS.304..391K}.
\emph{Kepler} will obtain exquisite
photometry on $\sim$100,000 stars, of which about 30 are expected to
host hot Jupiters with periods less than 3 days \citep{2008arXiv0804.1150B}. \emph{Kepler} has the potential to
measure the gravitational quadrupoles of very hot Jupiters though the
technique described below. If successful, this will constitute a major
step towards an understanding of the diversity of planetary interiors.

In Section 2, we describe the background theory that connects 
interior structure and orbital dynamics and explore which effects are most
important. Section 3 applies this theory to the observable changes in
the transit photometry, including full \emph{Kepler} simulated
light curves. We show in Section 4 that the signal due to the planetary
interior has a unique signature. Other methods for
inferring planetary interior properties are discussed in Section
5. The final section discusses the important conclusions of our work.

\section{Background Theory}

\subsection{Coordinate System and Notation}
The internal structure of very hot
Jupiters can be determined by observing changes in the planet's
orbit. These changes can be described in terms of two general types of
precession. Apsidal precession refers to rotation of the orbital
ellipse within the plane of the orbit. It is characterized by
circulation of the line of apsides, which lies along the major axis of the 
orbit. Nodal precession, on
the other hand, occurs out of the plane of the orbit and refers to the
orbit normal precessing about the total angular momentum vector of the
system. For typical very hot Jupiter systems with no other planets, apsidal
precession has a much stronger observable signal than nodal
precession (see Section \ref{obliquitysec}), so we focus our discussion on 
the simpler case of a fixed orbital plane.

As is typical for non-Keplerian orbits, the star-planet
orbit is described using osculating orbital elements that change in time. We
identify the plane of the sky as the reference plane and orient the
coordinate axes in the usual way such that the sky lies in the x-z plane
with the y-axis pointing at Earth. The intersection of the
orbital plane and the reference plane is called the line of nodes, but
without directly resolving the system, there is no way to determine
the orientation of the line of nodes with respect to astronomical
North; thus, the longitude of the ascending node, $\Omega$,
cannot be determined. Given this degeneracy, we simplify the
description by orienting the z-axis to lie within the plane spanned by the orbit
normal and the line-of-sight. The angle between the line of sight and the orbit normal
is $i$, the inclination. The x-axis is in the plane of the sky and is the reference
line from which the argument of periapse ($\omega$) is measured (in
the standard counter-clockwise sense). For this choice of coordinates,
the argument of periapse and longitude of periapse ($\varpi$) are
equivalent. Given this coordinate system, transit centers occur when
the planet crosses 
the y-z plane; this point lies $90^{\circ}$ past the reference x-axis,
and thus primary transits occur when the true anomaly, $f$, satisfies $f
_{tr}+\omega_{tr}\equiv90^{\circ}$, where the subscript $tr$ indicates the
value at transit center.\footnote{In elliptical orbits, if the inclination is not
  90$^{\circ}$, the photometric minima do not exactly coincide with the
  planetary conjunctions. See \citet{1959cbs..book.....K}, p. 388 and section \ref{transitshapes}
below.}

Throughout this paper, we refer to
parameters of the star (mass, radius, etc.) with
subscripts of ``$*$'' and parameters of the planet with
subscripts of ``$p$''. 
For evaluation of various equations, we will take as
fiducial values the mass ratio $M_p/M_* = \e{-3}$,
the radius ratio 
$R_p/R_* = 0.1$ (though some low density planets have radius ratios
greater than 1/6), and the semi-major axis in stellar radii
  $a/R_* = 6$, typical for very hot Jupiters, which we define as planets 
with semi-major axes $a \lesssim 0.025$ AU (see Table 1).\footnote{Throughout this work, we do
  not distinguish between $M_{tot}$ and $M_*$, since $M_p \ll M_*$.} In this definition, 
we deviate from \citet{2008arXiv0804.1150B}, who define very hot Jupiters as planets
with periods less than 3 days. These authors estimate that $\emph{Kepler}$ will find $\sim$30 such planets, 
of which $\sim$16 will be brighter than V=14 (T. Beatty, pers. comm.).
Since our definition is more stringent, our 
technique will be applicable to fewer \emph{Kepler} planets.

\subsection{Rotational and Tidal Potentials}

It is well known from classical
mechanics, that if stars and planets are considered to be purely spherical
masses, then they will obey a simple $r^{-2}$ force law and hence
execute closed elliptical orbits. Non-spherical mass effects are caused by the
application of external potential(s): the centrifugal potential of
spinning bodies causes rotational flattening and the tidal potential
of a nearby mass raises tidal bulges. Rotational and tidal bulges
create gravitational quadrupole fields ($r^{-3}$) that lead to
orbital precession. 

The complex subject of how planets\footnote{For clarity, in
  these sections we focus on the planetary shape, though the derivations 
are also valid for stars.} respond to
applied potentials is encapsulated in the so-called theory
of figures \citep{1978ppi..book.....Z}. As long as the distortions
are small, we can simplify 
the problem by ignoring the small interaction terms between the tidal
and rotational potentials; in this paper, we thus restrict ourselves
to the first order theory, where the two planetary responses simply
add. Even in the linear case,
the way the fluid planet responds depends on the full radial density
structure of the planet. The planetary response is conveniently
captured in a single variable $k_{2p}$, using the definition 
\be
\label{k2def}
V_2^{\rm ind}(R_p) \equiv k_{2p} V_2^{\rm app}(R_p)
\ee
where $k_{2p}$ is the Love number of the planet,
which is just a constant of proportionality between the applied second
degree potential field $V_2^{\rm app}$ and the resulting field that it
induces $V_2^{\rm ind}$ at the surface of the planet. Due to the
orthogonality of the Legendre polynomials used to express the gravity field, 
if the planet is responding
to a second degree harmonic field, then only the second degree
harmonic of the planet's gravity field is altered, to first-order. Thus,
$k_{2p}$ is a measure of how the redistribution of mass caused by the applied
potential actually affects the external gravity field of the
planet. In the stellar literature, the symbol $k_2$ is used for the
apsidal motion constant, which is half of the secular/fluid Love
number that we use throughout this paper \citep{1939MNRAS..99..451S}.

The Love number $k_2$ is an extremely useful parameterization, as it
hides the complex interactions of a planet and an
applied potential in just a single number. The process of calculating
$k_2$ of a fluid object (like stars and gas giants), from the interior density
distribution is fairly straightforward and outlined in several places
\citep[e.g.,][]{1939MNRAS..99..451S,1959cbs..book.....K}. Objects with most of their mass
near their cores, like stars, have very low $k_{2}$ values \citep[$\sim$0.03
for main sequence solar-like stars,][]{1995A&AS..109..441C} since the
distorted outer envelope has little mass and therefore little effect
on the gravity field. Planets have much flatter density
distributions, and thus distortions of their relatively more massive outer 
envelopes
greatly affect the gravity field. At the upper extreme lies a uniform
density sphere, which has $k_2 = 3/2$. In this way, {\it $k_2$ can be
  thought of as a measure of the level of central condensation of an
  object}, with stronger central condensation corresponding to smaller 
$k_2$.

By examining the variations in $k_2$ for giant planets within our own 
Solar
System, we can gain a feel for its expected values and how sensitive
it is to internal structure. The $n=1$ polytrope is commonly used to approximate
the density structure of (cold) gas giant planets; it has $k_2 \approx
0.52$ \citep{1959cbs..book.....K}. This can be compared to the value 
determined
from the gravity measurements of Jupiter, where $k_{2J} \simeq 0.49$. Even
though Jupiter may have a ~10 Earth mass core, it is small in
comparison to Jupiter's total mass, and thus it has minor effect
on the value of $k_2$. Saturn, on the other hand has a roughly 20
Earth mass core and is less than 1/3 of Jupiter's mass. 
As a result, the presence of
Saturn's core is easily seen in the value of its Love number $k_{2S}
\approx 0.32$. From this, we can see that planets with and without
significant cores differ in $k_{2p}$ by about
$\sim 0.1$. This can also be inferred from \citet{2003ApJ...588..545B} by
using the Darwin-Radau relation to convert the moment of inertia
factor to $k_2$. Furthermore, \citet{2001ApJ...548..466B} list the moment
of inertia factors of various planet models of HD 209458 b and $\tau$
Bootis b, which correspond to a range of $k_{2p}$ values from $\sim$0.1 to $\sim$0.6. 

Current methods for inferring the internal structures of
extra-solar planets combine measurements of the mass and radius
with a model to obtain estimates of the planet's implied composition
and core size. Unfortunately, these models require one to make
assumptions about the degree of differentiation, among other things 
\citep{Baraffe2008}. A
good measurement of $k_{2p}$, however, reveals important independent 
structure information, which can break the degeneracies between
bulk composition and the state of differentiation. Given
such a wide range of potential $k_{2p}$ values, even an
imprecise measurement of $k_{2p}$ will be extremely valuable for
understanding extra-solar planets. By measuring the $k_{2p}$ values for
extra-solar planets, we can also uncover constraints on the
density structure that are independent of the measurement of the 
planetary radius. This new information may
allow us to probe the unknown physics responsible for the currently
unexplained radius anomalies.

\subsubsection{Induced External Gravity Field}
The internal structures of planets in our own solar system are most
readily characterized by the zonal harmonics of the planet's gravity
field, i.e. $J_2$, $J_4$, etc. It is these high-order harmonics that are
directly measured by spacecraft flybys. To better understand the
connection between the two, we can relate the $k_2$
formulation to $J_2$ by writing out the expression for the induced
potential at the surface of the planet in Equation \ref{k2def} in
terms of the definition of $J_2$, yielding: 
$k_{2p} V_2^{\rm app}(R_p) = -J_2 \frac{GM_p}{R_p}P_2(\cos\theta)$, where 
$P_2$ is the usual Legendre polynomial and $\theta$ is the planetary 
co-latitude \citep{1999ssd..book.....M}. 
We can use this equation to obtain expressions for the $J_2$ field 
induced by both rotation and tides (discussed in more detail below). 
The relation relies on dimensionless
constants which compare the strength of the acceleration due to
gravity with that of the rotational and tidal potentials:
\be
\label{qparams}
q_r = \frac{\nu_p^2 R_p^3}{G M_p} \qquad {\rm and} \qquad q_t =
-3\left( \frac{R_p}{r} \right)^3 \left(\frac{M_*}{M_p}\right)
\ee
where $\nu_p$ is the angular spin frequency of the planet. For the case 
where the spin axis and tidal bulge axis are
perpendicular (i.e. zero obliquity), the relationship between $J_2$
and $k_2$ is, to first order:
\be
\label{J2perp}
J_{2} = \frac{k_2}{3}\left(q_r - \frac{q_t}{2} \right)
\ee
Note that $q_t$ is a function of the instantaneous orbital separation, $r$,
and is thus constantly changing in an eccentric orbit in response to the
changing tidal potential. Hence $J_2$ for eccentric extra-solar planets is a
complex function of time. This is why it is more sensible to analyze the orbital
precession in terms of $k_2$, which is a fixed intrinsic property of
the planet, rather than $J_2$.

As very hot Jupiters are expected to be synchronously locked (denoted by $s$)
with small eccentricities, it can easily be shown that
$q_t^s\approx-3q_r$, which simplifies equation \ref{J2perp} yielding: 
\be
J_{2p}^s \simeq \frac{5}{6}k_{2p} q_r \simeq \frac{5}{6} k_{2p} \left( \frac{M_*}{M_p}
\right) \left( \frac{R_p}{a} \right)^3
\ee
Using a moderate value of $k_{2p}=0.3$, the $J_2$ of very hot Jupiters
reaches as high as 5 $\x \e{-3}$, about half of the measured $J_2$ of  
Jupiter and
Saturn.

\subsection{Apsidal Precession}
The quadrupole field created by rotational and tidal potentials 
discussed above induces precession of the star-planet orbit. 
Both Jupiter and Saturn have rather significant quadrupoles,
dominated entirely by their sizeable
rotational bulges resulting from rapid rotation periods
of less than 10 hours. In contrast, very hot Jupiters are expected to be synchronously
rotating, and thus their spin periods are longer by a factor of a
few. Since the rotational bulge size goes as the square of the spin
frequency, very hot Jupiters should have rotational bulges that are at
least an order of magnitude smaller than Jupiter and Saturn, inducing
only tiny quadrupole fields. These extra-solar planets are extremely
close to their parent stars, however, with semi-major axes of only
$\sim6$ stellar radii. Very hot Jupiters are thus expected to have large tidal bulges which are
shown below to dominate the quadrupole field and resulting apsidal
precession.

\subsubsection{Precession Induced by Tidal Bulges}
The orbital effect of tidal bulges is complicated by their continuously changing size.
While tidal bulges always point directly\footnote{We can ignore the 
lag due to dissipation, which has an angle of only $Q_p^{-1}
  \lesssim 10^{-5}$ for giant planets \citep{1966Icar....5..375G,1999ssd..book.....M}.} 
at the tide-raising object, their size is a function of orbital
  distance. Since the height of the tidal bulge depends on the actual separation
between the objects, the second-order gravitational potential is
time-varying in eccentric orbits. Accounting for this dependence
(which cannot be captured by using a fixed $J_2$) is critical, as
illustrated by \citet{1939MNRAS..99..451S}. The dominant tidal
perturbation to the external gravity field of the planet, evaluated at
the position of the star, is a second-order potential:
\be
V_{tid}(r) = \frac{1}{2} k_2 GM_* R_p^5 r^{-6} 
\ee

The apsidal precession due to the tidal bulge, including the effect of
both the star and the planet is \citep{1939MNRAS..99..451S,2001ApJ...562.1012E}: 
\begin{eqnarray}
\label{tidalprec}
\dot{\omega}_{\rm tidal}  & = & \dot{\omega}_{\rm tidal,*} +
\dot{\omega}_{\rm tidal,p} \nonumber \\
& = &
\frac{15}{2} k_{2*} \left( \frac{R_*}{a}\right)^5 \frac{M_p}{M_*} f_2(e) n \nonumber \\
& + &
\frac{15}{2} k_{2p} \left( \frac{R_p}{a}\right)^5 \frac{M_*}{M_p} f_2(e) n 
\end{eqnarray}
where $n$ is the mean motion and $f_2(e)$ is an eccentricity function: 
\begin{eqnarray}
f_2(e) & = & (1-e^2)^{-5}(1 + \frac{3}{2} e^2 + \frac{1}{8} e^4) \nonumber \\
     & \approx & 1 + \frac{13}{2} e^2 + \frac{181}{8} e^4 + ... \label{f2e}
\end{eqnarray}
Note that the factor of 15 does not appear for stationary
rotational bulges, as detailed below, and comes through Lagrange's
Planetary Equations from the higher dependence on radial separation
$(r^{-6})$ in the tidal potential. For this reason, tidal bulges
are much more important in producing apsidal precession. 

Furthermore, the main factor of
importance to extra-solar planets is the 
mass ratio, which comes in because the height of the tide is
proportional to the mass of the tide-raising body. Consider the ratio
of the planetary and stellar effects: 
\be
\frac{\dot{\omega}_{\rm tidal,p}}{\dot{\omega}_{\rm tidal,*}} = 
\frac{k_{2p}}{k_{2*}} \left( \frac{R_p}{R_*} \right)^5 \left(
\frac{M_*}{M_p} \right)^2 \simeq 100
\ee
For tidal bulges, the apsidal motion due to the planet clearly dominates over
the contribution of the star. Even though the planet's radius is
smaller than the star's by a factor of ten, the star is so much more massive
than the planet that it raises a huge tidal bulge, which consequently
alters the star-planet orbit. The benefit provided by the inverse
square of the small mass ratio is compounded by the
order of magnitude increase in $k_2$ of the planet over the star.

\subsubsection{Precession Induced by Rotational Bulges}
The quadrupolar gravitational field due to the planetary rotational bulge, evaluated
at the star's position is:
\be
V_{\rm rot}(r) = \frac{1}{3} k_2 \nu_p^2 R_p^5 r^{-3} P_2(\cos \alpha_p)
\ee
where $\alpha_p$ is the planetary obliquity, the angle between the orbit normal
and the planetary spin axis. \citet{1939MNRAS..99..451S} assumes zero 
obliquity and calculates the secular
effect of this perturbation on the 
osculating Keplerian elements. This final result, including the effect of both the star and the planet
is\footnote{The full equation, including arbitrary obliquities, is
  given in \citet{1978ASSL...68.....K}, Equation V.3.18 
\citep[see also ][]{1939MNRAS..99..451S,2001ApJ...562.1012E}. Also recall that, unlike these authors, we
use the symbol $k_2$ to represent the Love number which is twice the
apsidal motion constant called $k_2$ in eclipsing binary literature.}: 
\begin{eqnarray}
\label{dorot}
\dot{\omega}_{\rm rot} & = & \dot{\omega}_{\rm rot,*} +
\dot{\omega}_{\rm rot,p} \nonumber \\
& = &
\frac{k_{2*}}{2} \left( \frac{R_*}{a} \right)^5 \frac{\nu_*^2 a^3}{GM_*}
g_2(e) n \nonumber \\ 
& + & 
\frac{k_{2p}}{2} \left( \frac{R_p}{a} \right)^5 \frac{\nu_p^2
  a^3}{GM_p} g_2(e) n 
\end{eqnarray}
where $g_2(e)$ is another eccentricity function: 
\be
g_2(e) = (1-e^2)^{-2} \approx 1 + 2e^2 + 3e^4 + ... \label{g2e}
\ee

Evaluating the importance of this effect requires an understanding
of the spin states of very hot Jupiters and their stars. 
The rotation and spin pole orientation of very hot Jupiters should be
tidally damped on timescales $\lesssim 1$ MYr 
\citep[e.g.,][]{2004ApJ...610..464D, Ferraz-Mello2008}. We therefore assume
that all planets have reached the psuedosynchronous rotation rate
derived by \citet{1981A&A....99..126H}. The rotation rate of the star is usually much slower
since the tidal stellar spin-up timescale is much longer than $\sim$1 GYr
\citep{2007ApJ...665..754F}.

If both the star and the planet were spinning synchronously, the
stellar and planetary rotational bulges would have comparable contributions to apsidal precession. 
However, since the tidal bulge of the planet is a much more
important effect, we find that even fast-spinning stars have a very weak contribution to apsidal
precession. 

\subsubsection{Total Apsidal Precession}

The other major contributor to the apsidal precession in extra-solar
planetary systems is general relativity. The anomalous apsidal
advance of Mercury's orbit due to its motion near the massive Sun was
one of the first confirmations of general 
relativity. This same apsidal advance is prevalent in
very hot Jupiter systems and has been shown to be possibly detectable
through long-term transit
timing \citep{2002ApJ...564.1019M,2007MNRAS.377.1511H,PK08,JB08}. The
relativistic advance is given (to lowest order) by: 
\be
\dot{\omega}_{GR} = \frac{3 G M_* n}{ac^2(1-e^2)}
\ee

One additional effect for non-synchronous planets is due to thermal tides
\citep{2009arXiv0901.0735A}, which create a bulge on the planet due to  
temperature-dependent expansion of an unevenly-radiated upper atmosphere. 
The thermal tidal bulge is very small in mass and is not expected to provide a 
significant contribution to apsidal precession (P. Arras, pers. comm.) 
and is thus neglected.

Since we are considering only the lowest-order effects, all the
apsidal precession rates (rotational/tidal for the star/planet and
general relativity) simply add to give the total apsidal
precession (roughly in order of importance for very hot Jupiters): 
\be
\dot{\omega}_{\rm tot} = \dot{\omega}_{\rm tid,p} + \dot{\omega}_{\rm GR}
 + \dot{\omega}_{\rm rot,p} + \dot{\omega}_{\rm rot,*} + \dot{\omega}_{\rm tid,*}
\ee
We are ignoring the small cross-terms (geodetic
precession, quadrupole-quadrupole coupling, Lense-Thirring effect,
nutation, etc.) for the purposes of this paper as higher-order corrections.

Calculating each of these contributions to the precession shows that
\emph{for very hot Jupiters, the dominant term in the total apsidal 
precession is due to the
  planetary tidal bulge}. For the known transiting planets, the
fraction of apsidal precession due to the planet is calculated and
illustrated in Figure 1. The precession due to the interiors of very
hot Jupiters towers over the other effects. General relativity, the
next largest effect 
is $\sim$10 times slower than the precession caused by the planetary
tidal bulge.  

The apsidal precession rate of very hot Jupiters due solely to the
interior structure of the planet is: 
\begin{eqnarray}
\dot{\omega}_{\rm p} & \approx & 3.26 \x \e{-10} \textrm{ rad/sec } \x \left( 
\frac{k_{2p}}{0.3} \right)
\left( \frac{M_*}{M_{\astrosun}} \right)^{3/2} \x \nonumber \\
& & \left( \frac{M_p}{M_{\rm J}} \right)^{-1} 
\left( \frac{R_p}{R_{\rm J}} \right)^{5}
\left( \frac{a}{0.025 \textrm{ AU}} \right)^{-13/2}
\end{eqnarray}
which explains why low density very close-in Jupiters are the prime targets for
measuring apsidal precession. For these planets, the precession rate can
reach a few degrees per year.

The precession due to the planet has generally been neglected in extra-solar
planet transit timing work 
to date \citep{2002ApJ...564.1019M,2007MNRAS.377.1511H}, which has
considered stellar oblateness or general relativity to be the dominant
effects (in the absence of other planets) though \citet{JB08} have
also pointed out that $\dot{\omega}_{\rm tidal,p}$ can be an
important source of apsidal precession. We find that the planetary
quadrupole is usually 1-2 orders of magnitude more important than effects
previously considered for single very hot Jupiters. Hence, measuring
apsidal precession essentially gives $\dot{\omega}_{\rm tid,p}$
which is directly proportional to 
$k_{2p}$, implying that transit light curve variations due to apsidal
precession can 
directly probe the interiors of extra-solar planets.

\begin{figure}						 
\begin{center}
\includegraphics[width=3.5in]{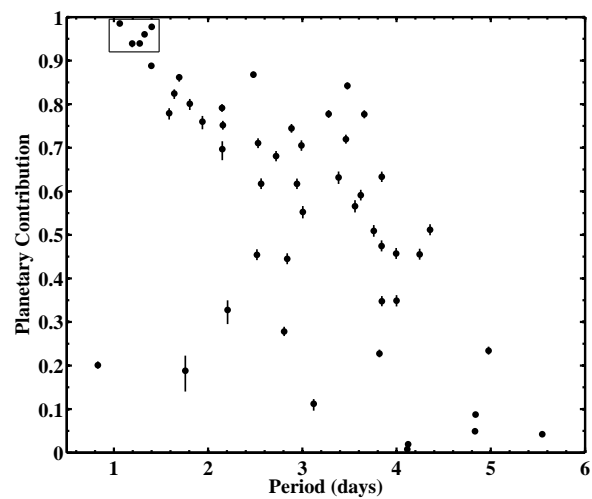}
	\caption{\label{odotfig} 
	  \textbf{Fraction of Apsidal Precession Due to the Planetary Quadrupole.}
  The points show the planetary fraction of
          the total apsidal precession calculated for the known transiting
          extra-solar planets with properties taken from J. Schneider's Extra-Solar Planet Encyclopedia
	  (\texttt{http://www.exoplanet.eu}), assuming the planet has a typical Love number
          of $k_{2p}=0.3$ (e.g. Saturn-like). The apsidal precession 
induced by the tidal and
	  rotational bulges of the planet overcome precession due to general relativity 
          and the star, especially for short period planets. 
	The "error bars" show the range of planetary
          contributions for a 5\% variation in stellar masses (and hence 
	$\dot{\omega}_{\rm GR}$) and the comparatively smaller effect of varying the stellar
          Love number and rotation rate over all reasonable values. 
	The five cases where
	  the planetary contribution to apsidal precession is most
	  important (boxed) also have the shortest precession periods:
	  WASP-12b, CoRoT-1b, OGLE-TR-56b, WASP-4b, and TrES-3b would
	  fully precess
	  in about 18, 71, 116, 120, and 171 years, respectively. The planet in the lower left is CoRoT-7, 
a super-Earth planet whose planetary contribution to precession is small because of its small radius. Transiting 
planets with periods longer than 6 days all had planetary contributions less than 0.15. 
	In all cases,
          the dominant signal in apsidal precession of very hot Jupiters is $k_{2p}$, which
          is determined by their internal density distribution and is a powerful
	  probe into their interior structure.  
         }
\end{center}
\end{figure}

\subsection{Modification of the Mean Motion}

Non-Keplerian potentials also modify the
mean-motion, $n$, and cause a 
small deviation from Kepler's Third Law. Including the effects
described above, the non-Keplerian mean motion, $n'$, 
is (dropping second-order corrections): 
\be 
n' = n \left( 1 + \epsilon - \frac{3GM_*}{2ac^2} \right)
\ee
where $\epsilon$ is defined as 
\begin{eqnarray}\label{epsilon}
\epsilon & = & \frac{k_{2*}}{2} q_{r,*} \left( \frac{R_*}{a}
\right)^2 + \frac{k_{2p}}{2} q_{r,p} \left( \frac{R_p}{a} \right)^2 \nonumber \\
& + & 3 k_{2*} \frac{M_p}{M_{tot}} \left( \frac{R_*}{a} \right)^5 + 
3 k_{2p} \frac{M_*}{M_p} \left( \frac{R_p}{a} \right)^5
\end{eqnarray}
and $n^2 \equiv \frac{GM_{tot}}{a^3}$. The general
relativistic correction to the mean motion is from \citet{1989racm.book.....S}. (Throughout
this paper, except where noted, the difference between $n'$ and $n$ is ignored as a higher-order
correction.)

As with apsidal precession, the planetary quadrupole is more important than the stellar
quadrupole by about 2 
orders of magnitude. At the largest, the correction to the mean motion
is a few times $\e{-5}$. \citet{2006NewA...11..490I} used the fact
that quadrupole moments cause deviations to Kepler's Third Law to
attempt to derive the $J_2$ of the star HD 209458 (the quadrupole of the 
planet was incorrectly ignored).

However, as \citet{2006NewA...11..490I} found, this method is only feasible if
you know the masses and semi-major axes of the orbit \emph{a priori}
or independently from Kepler's Law. Since the error in stellar masses
(from radial velocities and evolutionary codes) is usually 3-10 \%
\citep[e.g.][]{Torres2008}, the propagated error on 
$k_{2p}$ would be a few times greater
than the highest $k_{2p}$ expected, making this method
impractical. It has been proposed that the stellar mass and semi-major axis can be
precisely and independently measured via the light-travel time effect described by
\citet{2005ApJ...623L..45L}. In practice, however, the light-travel
time effect is highly degenerate with the unknown transit epoch and/or the
orbital eccentricity. We find that a precise independent measurement of $M_*$
from light-travel time is
impractical even 
with the excellent photometry of \emph{Kepler}.\footnote{We do note that 
detailed observations of multiple-planet systems can
yield mass estimates of each of the bodies independently. 
\emph{Kepler} asteroseismology can also provide independent information about
stellar mass and other properties \citep{2008arXiv0807.0508K}.}

\subsection{Expectations for Planetary Eccentricities}

Thus far, we have quantified how planetary interiors affect the orbit
through precession. The photometric observability of this apsidal
precession is highly dependent on the current orbital eccentricity
($e$). Small eccentricities are the
largest limitation to using transit light curves to probe extra-solar planet
interiors. Indeed, if eccentricities are very low, 
measuring apsidal precession from transit light curves may not be possible
for any of the \emph{Kepler} planets.

Nearly all hot Jupiters have eccentricities consistent with zero,
though the radial velocity technique has difficulty putting 3-$\sigma$ upper
limits on eccentricities smaller than 0.05 \citep{2005ApJ...629L.121L}. So far, the  
strongest constraints are placed by comparing the deviation of the
secondary transit time from half the orbital period, which are related
by \citep[e.g.,][]{2005ApJ...626..523C}:
\be \label{heqoffset}
e \cos \omega \simeq \frac{\pi}{2P_{\rm orb}} (t_{\rm sec}-t_{\rm
  prim}-\frac{P_{\rm orb}}{2})
\ee

Similarly, by measuring the primary and secondary transit durations
($\Theta_I$ and $\Theta_{II}$), an additional constraint can be placed on
$e \sin \omega$. The equation commonly quoted in the extra-solar
planet literature \citep{1999ebs..conf.....K,2003ASPC..294..449C,
2006ApJ...653L..69W}
has a sign error; the correct equation is derived by
\citet{1959cbs..book.....K}, p. 391 :
\be \label{keqdurratio}
e \sin \omega = \frac{\Theta_{II}-\Theta_I}{\Theta_{II}+\Theta_I} 
\frac{\alpha^2 - \cos^2 i}{\alpha^2 - 2 \cos^2 i}
\ee
where $\alpha \equiv \frac{R_*+R_p}{a\sqrt{1-e^2}}$. The accuracy of this measurement is typically smaller than for
$e \cos \omega$, but we include this equation to note that there is information 
about both the eccentricity and its orientation in the full transit light curve \citep[see also][]{2009arXiv0901.0282B}. \label{eccexpect}

Combining secondary transit timing information with radial velocity and
Rossiter-McLaughlin measurements to help constrain $\omega$,
\citet{2005ApJ...631.1215W} found the best-fit eccentricity for HD 209458
was $\sim$0.015. Though \citet{2005ApJ...631.1215W} argue that the actual eccentricity is  
probably less than 0.01, it is not necessarily 0 \citep{2007MNRAS.382.1768M}. 
Recently, \citet{2008arXiv0806.1478J} revealed WASP-14b, a young massive hot
Jupiter with an eccentricity of 0.1; WASP-10b and WASP-12b also appear to be
eccentric \citep{2008arXiv0806.1482C,2009ApJ...693.1920H}, though these
eccentricities may be spurious or overestimated. 

The most accurate eccentricity constraint is
a detection by \citet{2007Natur.447..183K} for the very hot Jupiter
HD189733b. They observed continuously and at high cadence (0.4
seconds) with the
Spitzer space telescope and measured a secondary timing offset
corresponding to $e \cos \omega = 0.001 \pm 0.0002$, a 5-$\sigma$
result that they could not explain by any other means. (Preliminary analysis
of additional data for this planet by \citet{2009IAUS..253..209A} 
indicates $e \cos \omega = 0.0002 \pm 0.0001$.) The constraint on $e \sin \omega$ is
much weaker. A non-zero
eccentricity of $e \simeq 0.003$ for hot Jupiters is therefore
consistent with every 
measurement available in the literature, though the actual values of eccentricities
at the $\e{-3}$ level are essentially unconstrained. 

In the absence of excitation, the current eccentricities of
these planets depend on 
the initial eccentricity and the rate of eccentricity
decay. Extrapolating from planets in our solar system \citep{1966Icar....5..375G}
  implies short circularization
  timescales of $\simeq$ 10 MYr, though recent studies have shown that using a fixed eccentricity
damping timescale is an inappropriate simplification of the full tidal evolution \citep[e.g.][]{Jackson2008,2009ApJ...692L...9L,2009arXiv0903.0763R}. 
Even an analysis using the full tidal evolution equations cannot give a compelling case
for the present-day eccentricities of these planets, since there are essentially no
  direct constraints on the tidal dissipation parameter for the planet, $Q_p$. Various estimates
show that $Q_p$ for exoplanets is not known and may be quite large \citep[e.g.,][]{2008ApJ...686L..29M}, 
implying that non-zero eccentricities are not impossible. Even so, we stress
that the best candidates for observing apsidal precession are also those 
planets that have the fastest eccentricity damping, since the damping timescale
and apsidal precession rates are both proportional to $\frac{M_p}{M_*} \left( \frac{a}{R_p} \right)^5$. 
Hence, those planets which have the fastest precession rates will also have the lowest
eccentricities. The first step in determining if this trade-off allows for apsidal precession to be measured by \emph{Kepler} data
is to apply the techniques described in this paper to the data themselves. 
Furthermore, with the discovery and long-term characterization of more planets using 
ground and space-based observations, the detectability of apsidal precession 
will increase dramatically.

We should note that there are several mechanisms that can excite eccentricities and
compete with or overwhelm tidal dissipation. The most prevalent is assumed to be
eccentricity pumping by an additional companion
\citep{1979Sci...203..892P,2001ApJ...548..466B,2006ApJ...649.1004A}. 
Even very small (Earth-mass or less) companions 
in certain orbits can provide significant eccentricity
excitation \citep{2007MNRAS.382.1768M}. (In this case, however, our
single-planet method for estimating $k_{2p}$ would need to be modified considerably.) Tidal dissipation 
in rapidly rotating stars tends to increase the eccentricity, potentially prolonging circularization
in some systems \citep{Ferraz-Mello2008}. Very distant inclined
companions (e.g. a planet orbiting a star in a misaligned binary star system) can
induce Kozai oscillations that impart very large eccentricities on secular timescales 
\citep[e.g.,][]{2007ApJ...669.1298F}. \citet{2009arXiv0901.0735A} proposed that thermal tides can significantly 
affect the orbital and rotational properties of extra-solar planets, though their
conclusions appear to be overestimated \citep{2009arXiv0901.3279G,2009arXiv0901.3401G}. 
Finally, recent (not necessarily primordial) dynamical instabilities in the planetary system can also be
responsible for generating eccentricity which simply hasn't damped away yet
\citep{2005Natur.434..873F,2005Natur.435..466G,2007astro.ph..3166C,2008ApJ...675.1538T}.
We, therefore, continue our analysis under the possibility that some very hot Jupiters
may have non-zero eccentricities.

\section{Transit Light Curves of Apsidal Precession}

Previous studies of transit light curve variability due to
non-Keplerian perturbations have focused almost
exclusively on transit timing. In contrast, we model 
the full photometric light curve in order to estimate the detectability of $k_{2p}$. 
This will automatically include the effect of changing transit
durations, which are very useful for detecting apsidal
precession \citep{PK08,JB08}. In addition, using full photometry can
provide a more direct and realistic estimate of the detectability of
$k_{2p}$. Of course, the drawback is additional computational cost,
though we found this to  
be manageable, requiring less than 20 seconds to generate the $\sim2$ million 
photometric measurements expected from \emph{Kepler}'s 1-minute cadence over 
3.5 years.

\subsection{Our Transit Light Curve Model}

Determining the photometric light curve of a transiting system requires knowing 
the relative positions of the star and the planet at all times. These can be 
calculated 
by describing the motion of the planet with time-varying osculating orbital 
elements. 
When describing the motion of the planet using instantaneous
orbital elements, it is usually customary to ignore the periodic terms
by averaging, as in \citet{1939MNRAS..99..451S}, and calculate only the secular terms. These 
small periodic terms describe how
the orbital elements change within a single orbit as a function of the true anomaly, $f$, 
due to the 
non-Keplerian potential. In precessing systems, the value of the true anomaly at central
transit, $f_{tr} \equiv 90^{\circ}-\omega_{tr}$, changes subtly from one transit to the 
next, inducing slow variations in the osculating orbital elements at transit.
Therefore, we include in our model the dominant 
periodic changes in orbital elements
as a function of orbital phase, using $M_{tr} \approx f_{tr}$ as an appropriate 
approximation for low eccentricities. Using a direct integration
(described in Section \ref{obliquitysec}), we
verified that ignoring these periodic variations can cause non-negligible systematic
errors in determining transit times. The periodic changes are 
derived from the same disturbing potentials used above. We follow the
method of \citet{1959AJ.....64..367K} for calculating osculating elements from mean elements,
and assume zero obliquity. The
correction is similar to the correction to the mean motion,
which is also applied in our model. 
The correction to the semi-major axis, eccentricity, longitude of periapse, and mean
anomaly are $a_{\rm osc} = a_{\rm mean} + \frac{2ae}{1-e^2} \epsilon \cos
M \approx 2ae \epsilon \cos M$, $e_{\rm osc} = e_{\rm mean} + \epsilon (1 - \cos M)$, 
$\omega_{\rm osc} = \omega_{\rm mean} + \frac{\epsilon}{e} \sin M$,
and $M_{\rm osc} = M_{\rm mean} - \frac{\epsilon}{e} \sin M$ where $\epsilon$ 
is defined in Equation \ref{epsilon}. General
relativistic periodic corrections are also added; these are taken 
from \citet{1989racm.book.....S}, page 92 (with
$\alpha=0$, $\beta=\gamma=1$). Using our direct integrator (described below), we
verified that these corrections reproduced the actual orbit to
sufficient accuracy for this analysis as long as $e \gg \epsilon \sim 
\e{-5}$. Other corrections
are higher order in small parameters and are ignored.  

Our model uses these corrected elements to generate astrocentric Cartesian 
coordinates for a specific system inclination and, for completeness, also 
includes the effect of 
light-travel time \citep{2005ApJ...623L..45L} though we concur with 
\citet{JB08} and \citet{PK08} that the 
light-travel time change due to $\dot{\omega}$ is unimportant. The positions 
are then translated to photometric light curves using the quadratic 
limb-darkening code\footnote{Available at 
\texttt{http://www.astro.washington.edu/
agol/transit.tar.gz}} 
described in \citet{2002ApJ...580L.171M}. \emph{Kepler} data will have enough signal-to-noise to justify
using non-linear limb darkening laws
\citep{2007ApJ...655..564K}, but we do not expect that this
simplification will significantly alter our conclusions. 

In addition, we include the photometry of the secondary eclipse. As
suggested by \citet{2007ApJ...667L.191L}, very hot 
Jupiters can reach temperatures exceeding 2000 K, where their blackbody
emission at optical wavelengths is detectable by \emph{Kepler}. 
This thermal emission is added to the
reflected light of the planet, which appears to be small based on the
low upper limit of the albedo of HD 209458b and TrES-3 measured by
\citet{2007arXiv0711.4111R} and \citet{2008AJ....136..267W}, respectively. 
We find that in \emph{Kepler}'s observing bandpass of 430-890 nm  
\citep{2006Ap&SS.304..391K}, thermal emission of
very hot Jupiters can dominate over the weak reflected light.
We estimate the depth
of the secondary eclipse ($d_{\rm sec}$) in our simulated 
\emph{Kepler} data by assuming that 1\% of the light is reflected and
the other 99\% absorbed and reemitted as processed thermal blackbody
emission from the entire planetary surface (day and night sides). To
be conservative and to account for unmodeled non-blackbody 
effects, we divide the resulting planet/star flux
ratio by 2 \citep{2008arXiv0807.1561H}; the resulting depth of around 
$2 \x \e{-4}$ is consistent with the lower values of 
\citet{2008arXiv0803.2523B}, the tentative measurement of the
thermal emission from CoRoT-2b \citep{2009IAUS..253...91A}, and the detection of 
secondary eclipse emission from OGLE-TR-56b \citep{2009A&A...493L..31S}. 
We note that the best candidates for
detecting $k_{2p}$
are those with small semi-major axes and large radii; these same
planets have relatively large $d_{\rm sec}$ values (Table 1). Secondary
eclipses are very useful for determining $e$ and $\omega$. We will
also find that they can be important for observing apsidal precession.\label{secdepth}

Our model generates accurate photometry for an extra-solar planet
undergoing apsidal precession. Several
other small photometric effects have been discussed in the literature,
which we do not include. Most of these effects are periodic (e.g. the
reflected light curve) and
therefore will not affect the long-term trend of precession. Care will need
to be taken to ensure that slow changes due to parallax and proper motion, which should
be quite small for relatively distant stars observed by \emph{Kepler}
\citep{2008arXiv0807.0008R,2007ApJ...661.1218S} or changes
in the stellar photosphere \citep{2008arXiv0807.0835L} are not significant.
Non-Gaussian astrophysical noise of 
the star and other systematic
noise should degrade the accuracy with 
which $k_{2p}$ can be measured compared to our ideal photometry. The long-term variability of
the star can be interpolated away or modeled \citep{2009A&A...493..193L}, though it is not 
clear how short-term variability will affect transit light curves at \emph{Kepler}'s level
of precision. On the other hand, complimentary observations (e.g., warm Spitzer, HST,
radial velocities, JWST, etc.) should only enhance our understanding of the
systems studied.

\subsection{Accuracy of $k_{2p}$ measurement}
With an accurate photometric model of apsidal precession, one could estimate the 
measurement accuracy of $k_{2p}$ 
from \emph{Kepler} data by carrying out a full Monte Carlo study of the inversion 
problem, going from realistic synthetic photometric data sets to a determination of all 
system parameters. In this work, instead, we carry out a much
simpler calculation which cannot provide strict one-sigma error estimates
like the Monte Carlo analysis, but does give an indication of
how well $k_{2p}$ can be resolved given a large dataset.

\begin{figure}						 
\begin{center}
\includegraphics[width=3.5in]{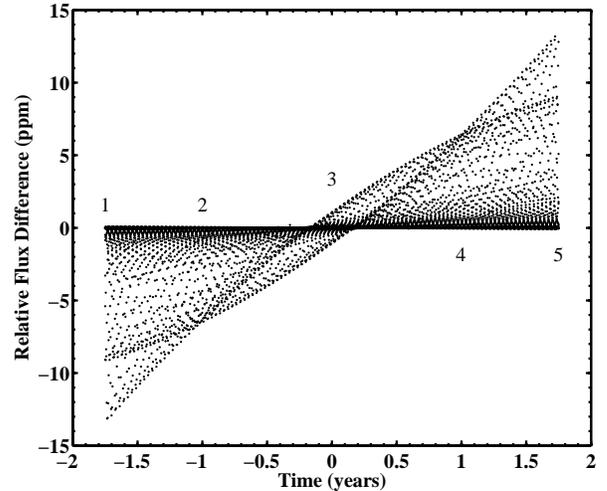}
	\caption{\label{candywrappermain} 
	  \textbf{Photometric Difference Signal from $k_{2p}$.}
      As described in the text, we use the difference between two
	  theoretical light curves in the
	  transit photometry to assess the observability of apsidal precession by
	  \emph{Kepler}. For WASP-4b at $\omega=0^{\circ}$, $e=0.003$, and a
	  central impact parameter, the difference between a model
	  with $k_{2p}=0$ and $k_{2p}=0.146$ would yield an effective
	  ``signal-to-noise'' of 1 on a moderately bright star
	  ($V=14$). Shown is this difference signal; the root sum of
	  squares of the signal is equal to 1000 ppm, the expected
	  photometric accuracy of \emph{Kepler} for a 1 minute
	  observation \citep{2006Ap&SS.304..391K}. The trends seen in the figure are
	  illustrated in Figure \ref{candywrapperpieces} 
	  by considering excepts of single primary transits from
	  the regions labeled 1-5. 
         }
\end{center}
\end{figure}

We obtain this accuracy estimate by comparing a realistic precessing
photometric model with $k_{2p} \ne 0$ to a base model with $k_{2p}= 0$. The
base model is still undergoing very slow apsidal precession, induced by
general relativity and $k_{2*}$. We
calculate the effect of a non-zero $k_{2p}$ value by subtracting the precessing 
model from the base model. (See Figures \ref{candywrappermain} and 
\ref{bowtiemain}.) Then, by calculating the root-sum-square of the residual
signal and comparing it to the photometric error on a single data
point, we obtain a numerical measure of the relative signal induced by
$k_{2p}$. The ``signal-to-noise'' ratio for the data set
is therefore given by:
\be
\frac{\rm S}{\rm N} \sim \frac{\sqrt{\sum_i (y_i -
    y_i^0)^2}}{\sigma}
\ee
where $y_i$ and $y_i^0$ are the photometry model values for the $k_{2p}$
test model and the base model, respectively, and $\sigma$ is the
photometric error. We use $\sigma = 1000$ parts per million (ppm) flux per 1-minute integration,
 corresponding to the expected noise of 
\emph{Kepler} on a faint $V=14$ star 
\citep{2006Ap&SS.304..391K}. Of the 30 planets with periods less than 3 days, 16 are expected
to be brighter than $V \simeq 14$ (T. Beatty, pers. comm.) and we can reasonably expect
some fraction of these to have 
orbits comparable to the planets modeled here.

Since our residual signal changes
as a function of time, this is not a true signal-to-noise calculation;
the distribution of values in time matters for a proper interpretation, but any 
distribution would
yield the same effective $\frac{S}{N}$, and thus this construction is
not capturing all of the details. Even so, it does provide a
useful and reasonable rough estimate for detectability. In order to
identify the resolution on the $k_{2p}$ measurement, we search for the
value of $k_{2p}$ which yields a signal-to-noise of $\frac{S}{N} =
1$. This is reasonable since it represents the threshold value of
$k_{2p}$, below which planetary induced precession cannot be distinguished
in the data with the given errors. The threshold $k_{2p}$ value can also be
loosely thought of as an estimate of the 1-$\sigma$ expected errors. \label{threshktwop}

\begin{figure}						 
\begin{center}
\includegraphics[width=3.5in]{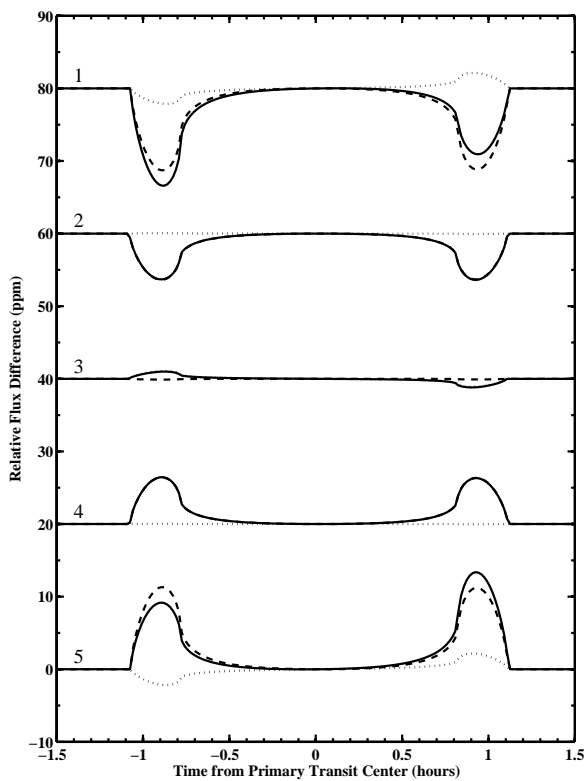}
	\caption{\label{candywrapperpieces} 
	  \textbf{Excerpts of Photometric Difference Signal.}
	  Examining excerpts of the residual signal shown fully in Figure
	  \ref{candywrappermain}, the effects of transit timing and
	  ``transit shaping'' can both be seen. The five excerpts are offset 
      for clarity.
	  Transit timing has an
	  asymmetric signal (dotted lines), obtained when
	  subtracting two transit curves 
	  slightly offset in time. Transit shaping, which is mostly due to changing
	  transit duration, creates a symmetric signal (dashed
	  lines). The total difference signal (solid lines) is
	  dominated by the effect of transit shaping, which has
	  $\sim$30 times more signal than transit timing alone. (See
	  explanation in text.) Both effects are maximized at the
	  beginning (1) and end (5), as
	  expected for a signal that increases with longer
	  baseline. The maximal signal occurs during ingress and
	  egress, when the light curve changes the fastest. The transit
	  shapes are equivalent at the center 
	  (3) by construction. The transit timing anomaly of precession is quadratic,
	  which, when fitted with a best-fit straight line
	  corresponding to a non-precessing signal, yields two
	  intersections when 
	  transit timing is minimized (2,4). The transit timing offset
	  at the beginning and end is only 0.085 seconds, while the
	  center is offset by -0.042 seconds. 
         }
\end{center}
\end{figure}

\begin{figure}						 
\begin{center}
\includegraphics[width=3.5in]{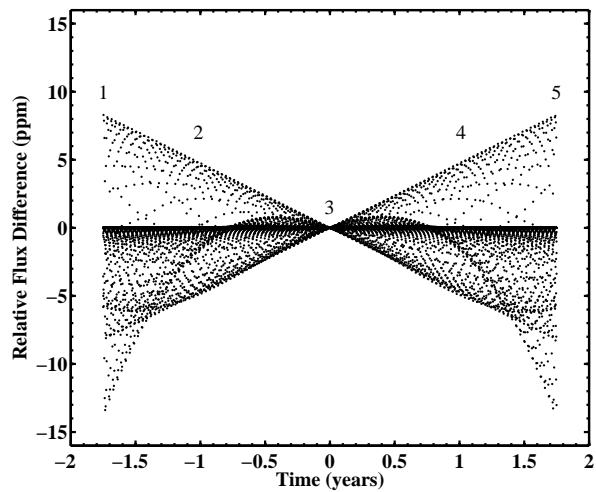}
	\caption{\label{bowtiemain} 
	  \textbf{Photometric Difference Signal from $k_{2p}$.}
	  Similar to Figure \ref{candywrappermain}, but for
	  $\omega=90^{\circ}$. This figure is dominated by the
	  photometric difference between secondary transits slightly offset in
	  time. At 
	  $\omega=90^{\circ}$ the changes in the primary
	  transits due to precession are small, except far away from the central time. 
      At this orientation, the primary-secondary timing offset 
	  (Equation \ref{heqoffset}) is
	  maximized. This ``secondary transit timing'' signal is
	  weaker than the signal from primary transit as the secondary
	  transit depth is much shallower.  Therefore, an unreasonably high $k_{2p}$ of 0.925 is required to detect the apsidal precession.
	  Excerpts of single secondary
	  transits taken from regions labeled 1-5 are shown in Figure
	  \ref{bowtiepieces}. 
         }
\end{center}
\end{figure}

\begin{figure}						 
\begin{center}
\includegraphics[width=3.5in]{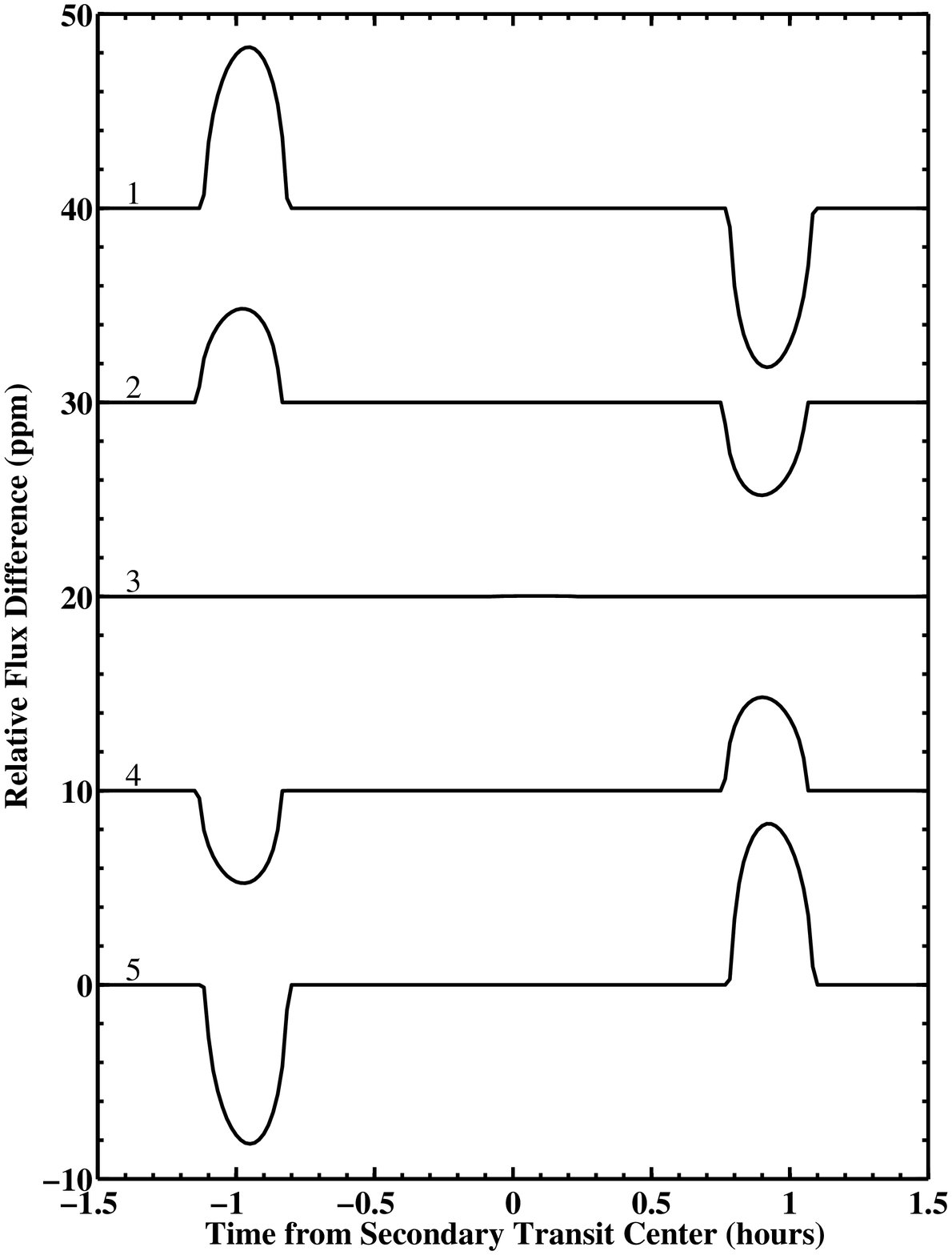}
	\caption{\label{bowtiepieces} 
	  \textbf{Excerpts of Photometric Difference Signal.}
	  Similar to Figure \ref{candywrapperpieces}, but for
	  $\omega=90^{\circ}$. Single secondary transit differences
	  are excised 
	  from the full difference signal shown in Figure
	  \ref{bowtiemain}. The shape of the curves is due to the
	  subtraction of two secondary transits slightly offset in
	  time. Since the secondary transits are complete
	  occultations, they are flat-bottomed and lack the additional
	  structure due to limb-darkening seen in Figure
	  \ref{candywrapperpieces}. By construction, the offset grows
	  in time away from the 
	  center (3) of the signal and attains a maximum at the
	  beginning (1) and end (5). Curves 2 and 4 are shown for
	  comparison to Figure \ref{candywrapperpieces}.
         }
\end{center}
\end{figure}

This is a realistic estimate only insofar as the residual signal
($y_i-y^0_i$) is due only to $k_{2p}$ and cannot be absorbed by any
other parameters. Hence we seek to choose other parameters so as to
minimize the residuals without changing $k_{2p}$. For most system
parameters, this is accomplished by referencing the time to the center
of the data set, and thus the difference between the signals grows
similarly forward and backward in time as seen in Figures
\ref{candywrappermain} - \ref{bowtiepieces}. The transit shapes in
both models are equivalent at the center of the dataset as would be
expected in an analysis of actual data.

Additionally, a major \label{degeneracies}
effect from changing the precession period is to alter the observed
average period. When analyzing actual data, this would just be
absorbed into a small adjustment to the (unknown) stellar mass, thereby
adjusting the period to absorb much of the $k_{2p}$ signal. It is therefore
important to correct for the average period change to avoid
significantly overestimating the signal due to $k_{2p}$. Additionally, 
there is a similar, though less severe, effect for the epoch of the
first transit, which is also adjusted to best absorb signal. This is
achieved by using an analytic expression for the transit times (see
Equation \ref{timeanombig} below) which match the transit times of the 
photometric model to very high accuracy. By fitting a line to these times, we
can determine the average period and epoch that absorb the
degenerate portions of the $k_{2p}$ signal, leaving behind the
residual due only to $k_{2p}$. We have not explicitly accounted for
degeneracies between the 
signal from $k_{2p}$ and the other parameters, like the radius, limb
darkening, and system inclination, but since $k_{2p}$ induces a time varying
signal while these other parameters are generally constant, there is
little expected signal absorption from these parameters. 

The only
major drawback of this approach is that it does not allow the
eccentricity state of the system to change. With real data, the
eccentricity and precession phase are not known in advance, and thus
must be found by inversion. As detailed in Section \ref{eccexpect}, 
eccentricity and orbital orientation are primarily constrained by
comparing primary and secondary transit pairs, and thus proper inversion is 
greatly aided by accurate observations in
wavelengths more favorable to secondary transit
observations, obtained by Spitzer, HST, or from the ground
 \citep[e.g.][]{2007Natur.447..183K,2008Natur.452..329S,2008arXiv0806.4911G}.
We also find that binned and folded \emph{Kepler} data has comparable sensitivity to 
a single Spitzer observation for characterizing the secondary eclipses of very hot Jupiters.
In any case, our assessment of the threshold $k_{2p}$ assumes that the eccentricity 
of the system is very well known, which will likely require additional supporting observations.

\subsection{Comparison to Expected Signal}

The residual light curves calculated for each planet, Figures
\ref{candywrappermain} - \ref{bowtiepieces}, match the theoretical
expectations of the apsidal precession signal \citep{2002ApJ...564.1019M,2007MNRAS.377.1511H,PK08,JB08}. 
To interpret the results of our 
analysis, it will be useful to briefly review
the major components of the apsidal precession signal: changes in the times of 
primary transits, changes in the shape of primary transits, and changes in the 
primary-secondary offset times \citep{2002ApJ...564.1019M,2007MNRAS.377.1511H,PK08,JB08}.

The primary transit times, $T_N$, due to apsidal precession are 
well described by a sinusoid for very low eccentricities ($e \ll 0.1$):
\be\label{timeanom}
T_N = T_0 + NP_{\rm obs}+\frac{eP_{\rm obs}}{\pi}(\cos \omega_{tr,N}-\cos\omega_{tr,0})]
\ee
where $T_0$ is the epoch of the first transit, ${\omega_{tr,N} \equiv \dot{\omega}(T_N-T_0) + \omega_{tr,0}}$ is the 
argument of periapse for the $N^{\rm th}$ transit, 
and $P_{\rm obs}$ is the \emph{observed} period between successive transits, which deviates
from the actual orbital period since the orbit has precessed a small
amount between transits \citep{1973bmss.book.....B}. For small eccentricities, 
the amplitude of the transit timing variations due to $k_{2p}$ is:  
\be\label{timingamp}
\frac{eP_{\rm obs}}{\pi} \simeq 119 \textrm{ sec} \x \left( \frac{e}{0.003} \right)  \left(
\frac{a}{0.025 \textrm{ AU}} \right)^{3/2} \left(
\frac{M_*}{M_{\astrosun}} \right)^{-1/2}
\ee
Given that individual transit times can be measured with accuracies of only a few seconds,
even tiny eccentricities $e \lesssim \e{-5}$ can induce detectable transit timing variations on precessional 
timescales ($\sim\dot{\omega}^{-1}$).

For our analysis, we extended Equation \ref{timeanom} to fifth order in eccentricity allowing accurate determination of transit
times for eccentricities up to of order 0.1. We also require a correction for the effect of a non-central impact
parameter ($i < 90^{\circ}$, $e > 0$). For an inclined eccentric orbit, the apparent path of the planet across the 
stellar disk is curved. At orientations where the line
of sight is not along the major axis of the ellipse, the curved path is also asymmetric. Therefore, the times of photometric minima, $T_N$,
do not correspond exactly to the times of conjunction (when the planet crosses the $y-z$ plane and $f_{tr} \equiv 90^{\circ} -\omega_{tr}$).
We follow the correction from Equation VI.9-21 of \citet{1959cbs..book.....K}, who find that at photometric
minimum, $f_{tr} = 90^{\circ} - \omega'_{tr}$, where ${\omega'_{tr} \equiv 
\omega_{tr} + e \cos \omega_{tr} \cot^2(i) (1 - e \sin \omega_{tr} \csc^2(i))}$; in this corrective term, it is only required to keep terms up to second order in
eccentricity. Assuming that $i$ and $\dot{\omega}$ 
are constant, it can be shown that
\begin{eqnarray}\label{timeanombig}
T_N & = & T_0 + NP_{\rm obs} \nonumber \\
& + & \frac{P_{\rm obs}}{\pi} \Bigg[ e(\cos \omega'_{tr,N} - \cos \omega'_{tr,0}) \nonumber \\
& + & \frac{3}{8} e^2 (\sin 2\omega'_{tr,N} - \sin 2\omega'_{tr,0}) \nonumber \\
& + & \frac{1}{6} e^3 (\cos 3\omega'_{tr,N} - \cos 3\omega'_{tr,0})  \nonumber \\
& + & e^4 \Big( \frac{1}{16} (\sin 2\omega'_{tr,N} - \sin 2\omega'_{tr,0}) \nonumber \\
& - & \frac{5}{64} (\sin 4\omega'_{tr,N} - \sin 4\omega'_{tr,0}) \Big)  \nonumber \\
& + & e^5 \Big( \frac{1}{16} (\cos 3\omega'_{tr,N} - \cos 3\omega'_{tr,0}) \nonumber \\
& - & \frac{3}{80} (\cos 5\omega'_{tr,N} - \cos 5\omega'_{tr,0}) \Big) \Bigg] \nonumber \\
\end{eqnarray}
This transcendental equation is solved iteratively for ($T_N - T_0$) to obtain the transit times and has been tested thoroughly against the empirical
determination of transit times calculated by our light curve model described above. 

The expected apsidal precession periods
(including small contributions from GR and the star) for
 WASP-12b, CoRoT-1b, OGLE-TR-56b, WASP-4b, and TrES-3b are around 
18, 71, 116, 120, and 171 years, respectively. In other words, they have precession rates induced
by the planetary tidal bulge of a few degrees per
year, compared to a few degrees per century as the fastest general
relativistic precession \citep{JB08}. We caution that if $\frac{R_p}{a}$ for WASP-12b is
overestimated due to imprecise data \citep[e.g.][]{2007AJ....134.1707W}, then
the precession period would increase accordingly.

Even with such fast precession rates, the duration of observations will generally be much shorter than the
precession period. In addition, as discussed above, the linear timing anomalies 
will be absorbed into the effective period as a small change in the unknown stellar mass \citep{2007MNRAS.377.1511H,PK08,JB08}.
Therefore, detection of apsidal precession from primary transit times alone will
require a significant detection of the curvature over a small portion of a long-period
sinusoid. Since the curvature in Equation \ref{timeanom} is maximal at 
$\omega \approx 0,180^{\circ}$, these orientations have the best primary 
transit timing signal. Even at these orientations, detecting $k_{2p}$ from primary transit
times alone is difficult, since it can be shown that the signal strength is proportional to $e 
\dot{\omega}^2$, due to the need to detect curvature \citep{2007MNRAS.377.1511H}. 

When the observational baseline is much shorter 
than the decades-long precession period, utilizing the changing shape of the 
transits can significantly improve detectability of apsidal precession \citep{PK08,JB08}. 
Transit shapes are primarily determined by the orbital speed at transit $\dot{f}_{tr}$ and impact parameter
$b$, both of which depend on the precession phase $\omega_{tr}$. 
For small eccentricities, the orbital angular speed at transit is given simply by 
$\dot{f}_{tr} \simeq n(1+2e \cos \omega_{tr})$.
Changes in the impact parameter are somewhat more subtle, since $b$ is given by 
$r_{tr} \cos i / R_*$, where $r_{tr} \simeq a(1-e^2)/(1 + e \sin \omega_{tr})$ is the star-planet separation. Hence, the apparent
impact parameter of the planet can change for non-central transits, even when the orbital plane
remains fixed. The evolving transit shape of precessing orbits is determined by variations 
in both orbital speed and impact parameter. Simplifying the effect of transit shape by considering only the variations in transit duration as a 
function of $\omega_{tr}$, \citet{PK08} and \citet{JB08} find that these two effects are of comparable magnitude. 
These authors also show analytically that the two effects exactly cancel when $b=1/\sqrt{2}$. At this
impact parameter, the transit duration stays constant throughout apsidal precession. The full photometric
transit shape, however, still changes detectably in a precessing orbit, though the magnitude of 
signal is reduced (Figure \ref{k2refangfig}). 

The expected effect of changing transit shapes is fully consistent with 
the photometric difference signals calculated by our model (Figures \ref{candywrappermain} and \ref{candywrapperpieces}). 
Indeed, our model shows that transit shaping dominates the signal by a factor of 
$\gtrsim$30 (Figure \ref{candywrapperpieces}). We can also see that changes in the transit shape are maximized at orientations near 
$\omega \approx 0,180^{\circ}$ (as expected from Equation \ref{keqdurratio}). \label{transitshapes} 

For small eccentricities, the transit shaping 
signal strength is given by $\frac{S}{N} \propto e \dot{\omega} \propto e k_{2p}$. 
Therefore, when transit shaping dominates the observable signal, we should find that 
searching for the threshold $k_{2p}$ value that yields $\frac{S}{N} = 1$ results
in a power law relationship between threshold $k_{2p}$ and $e$, such that $k_{2p} \propto e^{-1}$.
By solving for threshold $k_{2p}$ for eccentricities from 0.001 to 0.1, we 
find, as expected, that threshold $k_{2p}$ very
closely follows a power law in eccentricity with a slope of -1 for all planets. This power law
relationship can be written as $ek_{2p} = C$, where $C$ is a constant calculated from our model that
depends on the planetary, orbital, and stellar parameters of the system.

At $\omega \approx 90,270^{\circ}$, transit timing and transit shaping effects are much weaker and 
are rather ineffective at constraining apsidal precession. 
At these orientations (when the Earth's line of sight is nearly aligned with the major axis of the orbit), 
another photometric signal emerges: variations in the difference between the times
of primary and secondary transits. The changing
orientation of the orbital ellipse causes a variation in the offset
between primary and secondary transit times following
Equation \ref{heqoffset} above \citep{2007MNRAS.377.1511H,JB08}. These authors show that
the strength of this signal is also proportional to $e \dot{\omega}$ and we find that the variation
in threshold $k_{2p}$ then also follows $k_{2p} \propto e^{-1}$.

\begin{figure*}						 
\begin{center}
\includegraphics[width=6in]{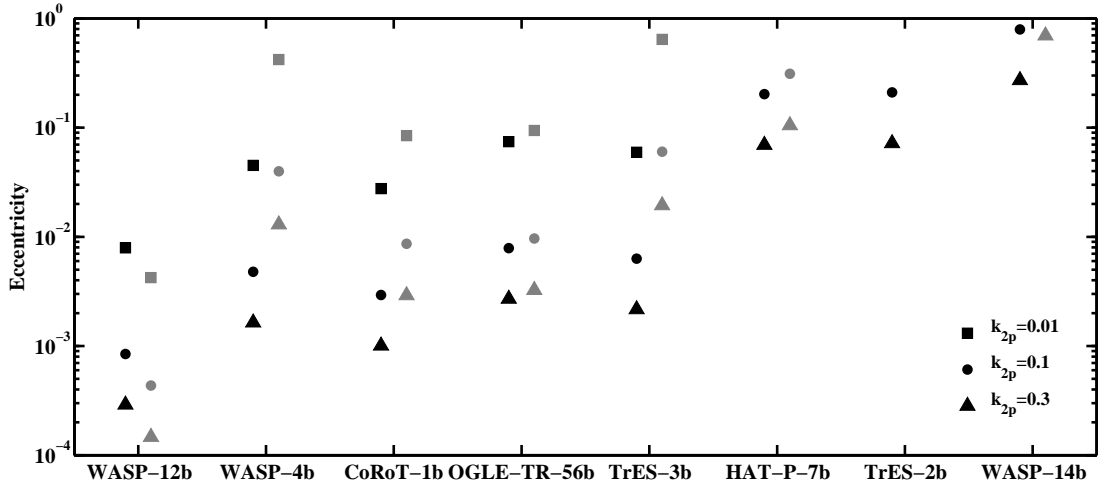}
	\caption{\label{ek2planet} 
	  \textbf{Eccentricities Needed to Detect Interior Properties
	  from Apsidal Precession.}
      The best-known planets for detecting $k_{2p}$
	  precession are analogs to the hot Jupiters WASP-12b, WASP-4b, CoRoT-1b, 
      OGLE-TR-56b, TrES-3b, HAT-P-7b, TrES-2b, and WASP-14b. Assuming that 
      analogs to these planets exist in the \emph{Kepler} field around a
      V=14 magnitude star, the above graph shows the eccentricities
      required to detect $k_{2p}$. Black
	  symbols correspond to calculations with $\omega=0^{\circ}$ and gray symbols
	  correspond to $\omega=90^{\circ}$; in both cases, $b=0$. Apsidal precession is much easier to detect
	  for larger eccentricities so increasing $e$ decreases the
	  detectable $k_{2p}$. Using our transit light curve model, 
      we found that threshold $k_{2p}$ values followed a power law
      $k_{2p} \propto e^{-1}$ (for low eccentricities), which is consistent 
      with the analytical estimates that $\frac{S}{N} \propto e\dot{\omega} \propto
      ek_{2p}$ (see Section \ref{threshktwop}). Interpolating (and sometimes
      extrapolating) on this power law relationship, the graph identified the
      eccentricities required of these analog planets to detect precession 
      due to a ``typical'' planetary interior of $k_{2p}=0.3$ (triangles). 
      For example, when $e=0.00026$ and $\omega=0^{\circ}$,
	  the apsidal precession due to an analog of WASP-12b should be just
	  detectable by \emph{Kepler}. A higher eccentricity (shown in Table 1) would be
	  needed to measure $k_{2p}$ with sufficient accuracy (0.1) to
	  distinguish between a massive core and a core-less model
	  (circles). Systematic errors are expected to become
	  important once the measurement error on $k_{2p}$ reaches as
	  low as 0.01 (squares). If any of the very hot
	  Jupiters discovered by \emph{Kepler} have
	  comparable eccentricities, the long-term high-precision
	  photometry would allow for a powerful probe into their interior structure. HAT-P-7b and TrES-2b
      are known to lie in the \emph{Kepler} observing field, but the values
      above are not corrected for improved photometric accuracy obtainable on these bright stars. Note that
	  the eccentricities shown above and in Table 1 are computed for $\frac{S}{N}=1$; 3-$\sigma$ measurements
      require eccentricities 3 times as high.
      }
\end{center}
\end{figure*}

\begin{figure}						 
\begin{center}
\includegraphics[width=3.5in]{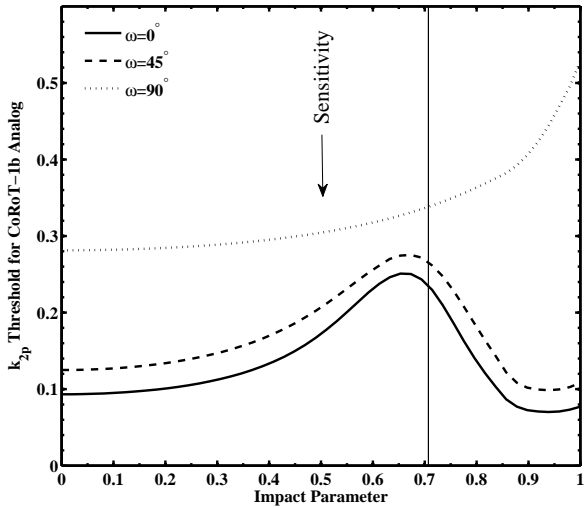}
	\caption{\label{k2refangfig} 
	  \textbf{Effect of Impact Parameter on Precession Signal.}
      The detectability of apsidal precession depends on the
	  impact parameter ($b$) of the orbital track across the
	  star. For $\omega=0^{\circ}$ (solid), the
	  signal of primary 
	  transits are most important, with transit shaping playing
	  the largest role. (See Figure
	  \ref{candywrapperpieces}.) However,
	  the strength of transit shaping is a function of impact
	  parameter with the minimum effect analytically estimated by \citet{JB08} and \citet{PK08} to be
	  $b=1/\sqrt{2}$ (vertical solid line). Using a full photometric
	  model, we see the expected decrease in the shaping signal
	  (i.e. requiring a larger $k_{2p}$ to reach
	  $\frac{S}{N}=1$). Note that the signal is nearly maximal, with small threshold $k_{2p}$ values, for a large
	  range of impact parameters. When $\omega=90^{\circ}$ (dotted), the
	  effect of primary transits are minimal and the offset in
	  secondary transits become the determining factor. (See
	  Figure \ref{bowtiemain}.) At high
	  impact parameters secondary eclipses are grazing, reducing
	  the observable signal. We also show the threshold $k_{2p}$ for an orientation of
      $\omega=45^{\circ}$, which lies, as expected, between the two extremes. The values 
      of threshold $k_{2p}$ shown are for an V=14 CoRoT-1b analog in the \emph{Kepler} field
	  with an eccentricity of 0.003. 
         }
\end{center}
\end{figure}

The photometric difference signal at $\omega=90^{\circ}$ is shown in Figures 
\ref{bowtiemain} and \ref{bowtiepieces}. Using the method described in Section \ref{degeneracies} to remove degeneracies 
almost eliminates the primary transit signal entirely, as expected, and the secondary 
transit offset becomes the more powerful signal. For WASP-12b, with an expected 
\emph{Kepler} secondary transit depth of $\sim$1830 ppm, the 
threshold $k_{2p}$ is actually \emph{lower} at $\omega=90^{\circ}$ (Figure 
\ref{ek2planet}). For the other planets, the secondaries are not as important. 

Our estimates of threshold $k_{2p}$ at $\omega=90^{\circ}$ are based on the unknown secondary transit depth ($d_{\rm sec}$) in the 
\emph{Kepler} bandpass (though our estimates of $d_{\rm sec}$ are consistent with all the measurements
in the literature to date). Furthermore, we find that $\frac{S}{N} \propto d_{\rm sec}$, so that deeper secondary
transits improve the accuracy with which $k_{2p}$ can be measured. It is important to note that combining \emph{Kepler} 
primary transit times with precise secondary transit times
measured in the near-infrared (e.g. by warm Spitzer, HST, or JWST) is a very powerful way to constrain apsidal
precession \citep{2007MNRAS.377.1511H} for any orientation. Even a few high-precision secondary eclipse observations 
are enough to lower the value of threshold $k_{2p}$ from our predictions, especially when $\omega \approx 90,270^{\circ}$.

By construction, threshold $k_{2p}$ values vary linearly with the 
assumed photometric error $\sigma = 0.001 \x 10^{0.2(V-14)}$. In addition, 
re-performing our analysis using a 6-year long \emph{Kepler} 
mission improved threshold $k_{2p}$ values by a common factor of $\sim$2.2. 

\begin{deluxetable*}{lrrrrrrrccccl}
\label{paramstable}
\tabletypesize{\scriptsize}
\tablecaption{Extra-Solar System Parameters and Results}

\tablehead{
\colhead{Planet Analog} & \colhead{$M_*$} & \colhead{$R_*$} &
  \colhead{$M_p$} & \colhead{$R_p$} & \colhead{$a$} & 
  \colhead{$d_{\rm sec}$\tablenotemark{b}} & \colhead{$\dot{\omega}_{tot}$} & 
  \multicolumn{2}{c}{$e$ (Threshold $k_{2p}$=0.1)\tablenotemark{c} } &
  \colhead{Threshold $\dot{P}$\tablenotemark{d}} & \colhead{Threshold $Q_*$\tablenotemark{d}} & \colhead{Ref} \\
  \colhead{} & \colhead{$M_{\astrosun}$} & \colhead{$R_{\astrosun}$} & 
  \colhead{$M_{J}$} & \colhead{$R_{J}$\tablenotemark{a}} & 
  \colhead{AU} & \colhead{ppm} & \colhead{$^{\circ}$/yr} &
  \colhead{$\omega=0^{\circ}$} & \colhead{$\omega=90^{\circ}$} & 
  \colhead{ms/yr} & & }

\startdata

  WASP-12b & 1.35 & 1.57 & 1.41 & 1.79 & 0.0229 & 1830 & 19.9   & 0.0008 & 0.0004 & 0.95  & 92700 & 1 \\
  CoRoT-1b & 0.95 & 1.11 & 1.03 & 1.55 & 0.0245 &  314 & 4.96   & 0.0028 & 0.0085 & 0.93  & 12500 & 2,3 \\
 WASP-4b & 0.92 & 0.91 & 1.24 & 1.36 & 0.0234 &  109 & 2.91     & 0.0047 & 0.0394 & 0.68  & 9900 & 4 \\
   TrES-3b & 0.93 & 0.83 & 1.91 & 1.34 & 0.0228 &  106 & 2.04   & 0.0062 & 0.0614 & 0.53 & 13700 & 5 \\
 OGLE-TR-56b & 1.17 & 1.32 & 1.29 & 1.30 & 0.0236 &  451 & 3.00 & 0.0077 & 0.0096 & 1.36  & 24700 & 6 \\
 HAT-P-7 b & 1.47 & 1.84 & 1.77 & 1.36 & 0.0377 &  176 & 0.25   & 0.2085 & 0.3146 & 6.73 & 2800 & 7 \\
  TrES-2 b & 0.98 & 1.00 & 1.19 & 1.22 & 0.0367 &   18 & 0.13 & 0.2102 & \nodata\tablenotemark{e} & 2.94 & 350 & 8 \\
  WASP-14b & 1.21 & 1.31 & 7.34 & 1.28 & 0.0360 &  144 & 0.09   & 0.8352\tablenotemark{e} & \nodata\tablenotemark{e} & 3.92 & 5400 & 9 \\
    XO-3 b & 1.21 & 1.37 & 11.8 & 1.22 & 0.0454 &   46 & 0.04   & \nodata\tablenotemark{e} & \nodata\tablenotemark{e} & 8.00 & 1700 & 10 \\
 HAT-P-11b & 0.81 & 0.75 & 0.081 & 0.42 & 0.0530 & 0.2 & 0.01 & \nodata\tablenotemark{e} & \nodata\tablenotemark{e} & 29.2 & 0.1 & 11 \\
  CoRoT-7b & 0.91 & 1.02 & 0.028 & 0.16 & 0.0170 &    8 & 0.29 & \nodata\tablenotemark{e} & \nodata\tablenotemark{e} & 16.8 & 80 & 12 \\

\enddata

\tablerefs{
(1) \citet{2009ApJ...693.1920H} \quad 
(2) \citet{2009arXiv0903.1845B} \quad 
(3) \citet{Barge2008} \quad
(4) \citet{2009AJ....137.3826W} \quad
(5) \citet{2009ApJ...691.1145S} \quad
(6) \citet{2007AA...465.1069P} \quad
(7) \citet{2009IAUS..253..428P} \quad
(8) \citet{2007ApJ...664.1185H} \quad
(9) \citet{2009MNRAS.392.1532J} \quad
(10) \citet{2008ApJ...677..657J} \quad
(11) \citet{2009arXiv0901.0282B} \quad
(12) \texttt{www.exoplanet.eu}\tablenotemark{f}
}

\tablecomments{These system parameters were used to estimate the detectability of apsidal precession for these very hot Jupiter systems. The derivation 
of the values in the remaining columns is described in the text and in the footnotes below. For all systems, $k_{2*} = 0.03$ and quadratic limb 
darkening parameters $u_1=0.35$ and $u_2=0.4$ (appropriate for \emph{Kepler}'s bandpass) were used \citep{2002ApJ...580L.171M}. For reference, the 
measured eccentricity of WASP-12b, WASP-14b, HAT-P-11b, and XO-3b are 0.049 $\pm$ 0.015, 0.091 $\pm$ 0.003, 0.198 $\pm$ 0.046, and 0.2884 $\pm$ 0.0035 respectively. Other planets 
have unmeasured eccentricities or eccentricity upper limits of $\lesssim$0.05. A discussion of these results is provided in Section \ref{resultsdisc}.}

\tablenotetext{a}{We use $R_{J}\equiv71492$ km, the equatorial radius at 1 bar.}
\tablenotetext{b}{The estimated depth of the secondary transit in \emph{Kepler}'s bandpass (see Section \ref{secdepth}).}
\tablenotetext{c}{The eccentricity required (at two different values
of $\omega$) so that a $k_{2p}$ difference of 0.1 has an effective
signal-to-noise of 1 in all of \emph{Kepler} data for a V=14 star, corresponding
to a photometric accuracy of 1000 ppm/min. If analogs to these planets were
found by \emph{Kepler} with the given eccentricities, the internal density distribution would 
be measured well enough to detect the presence of a large core (see Section \ref{threshktwop}). These values 
correspond to the circles in Figure \ref{ek2planet}. These results are for central transits (for $b > 0$, see Figure \ref{k2refangfig}). }
\tablenotetext{d}{The value of the change in period, $\dot{P}$, that can be detected with a signal-to-noise of 1 in all of \emph{Kepler} data for a V=14 star (see Section \ref{threshpdot}). The value of threshold $Q_*$ is an estimate of the maximum value of the stellar tidal dissipation parameter, $Q_*$, assuming that the period decay is due entirely to tidal evolution 
of the planet. Lower values of $Q_*$ are detectable by \emph{Kepler}. Stars are thought to have time-averaged $Q_*$ values around 10000, though this value is highly uncertain and could be much higher for individual stars.}
\tablenotetext{e}{Even with the precision of \emph{Kepler}, apsidal precession for these planets
is undetectable. The extrapolation used to compute eccentricities at specific values of threshold $k_{2p}$ assumes the inverse relationship discussed in the text $k_{2p} \propto e^{-1}$, which is only true for low eccentricities.}
\tablenotetext{f}{This ultra-short period low-mass planet was recently announced by the CoRoT team, but has not been published in a peer-reviewed
journal. We take the parameters from J. Schneider's Extra-solar Planets Encyclopedia and use the mass-radius relation for terrestrial super-Earths of \citet{2007Icar..191..337S}
to estimate the mass as $\sim$9 Earth masses (rather than using the quoted upper limit of 17 Earth masses).}

\end{deluxetable*}

\subsection{Results for Specific Planets} \label{resultsdisc}
Using the method described above, we have determined the threshold
$k_{2p}$ for the most favorable known transiting planets as analogs for the 
very hot Jupiters to be discovered by \emph{Kepler}. The threshold
$k_{2p}$ for each planet was computed at a range of eccentricities
from 0.001 to 0.1 and for $\omega=0^{\circ}$ and
$\omega=90^{\circ}$. Using the relationship discussed above ($k_{2p} \propto e^{-1}$)
we interpolated (and sometimes extrapolated) our calculations to determine
the eccentricity required to reach threshold $k_{2p}$ values of 0.3, 0.1, and 0.01. 
These results are summarized in Figure \ref{ek2planet} and Table 1.

WASP-12b is the best candidate for observing apsidal precession. With an eccentricity of
$e \simeq 0.00026$ and $k_{2p}$=0.3, the apsidal precession would have
an effective signal-to-noise of $\sim$1 for all of \emph{Kepler}
data. If $e$ is $\sim$0.001, then $k_{2p}$ can be well
characterized and not just detected. As the difference in $k_2$
between Jupiter and Saturn of $\sim0.15$ is primarily due to the
presence of a massive core, a resolution in $k_{2p}$ of 0.1 is enough
to detect whether or not the planet has a core, at the $\sim$1-sigma level. 

Although WASP-12b does not lie in the \emph{Kepler} field, it 
clearly stands out as an excellent candidate for observing apsidal precession. Though the putative eccentricity of 
0.049 \citep{2009ApJ...693.1920H} is probably 
an overestimate \citep{2005ApJ...629L.121L}, if it were real, it would cause sinusoidal 
transit timing deviations with an amplitude of $\sim$25 minutes (using Equation \ref{timeanom}) and 
a period of $\sim$18 years. Such a large deviation would be readily observed from the ground in 
either transit times or transit shapes. If apsidal precession is not observed, tight upper limits
on the eccentricity can be established. 

Analogs to the very hot Jupiters WASP-4b, TrES-3b, CoRoT-1b, and OGLE-TR-56b are good candidates
for observing apsidal precession if the eccentricities are above $\sim$0.003. (Note that CoRoT-1b has
only $\sim$30 days of observations from the \emph{CoRoT} satellite \citep{Barge2008}, which is insufficient to observe
any of the effects discussed in this paper.) These planets have precession periods of around 100 years so that the argument
of periapse of these planets changes by $\sim$10$^{\circ}$ during the course of \emph{Kepler} observations. Though none 
of these planets lie in the \emph{Kepler} field, they are all good candidates for observing apsidal precession though precision
photometry.

WASP-14b is more massive and has a larger semi-major axis (0.035 instead of 0.025)
which is enough to significantly reduce the detectability of apsidal precession which only proceeds at 0.1$^{\circ}$ per year. 
Unlike the previously mentioned planets, WASP-14b has a known non-zero eccentricity of 0.091 $\pm$ 0.003 \citep{2009MNRAS.392.1532J}. 
Thus, the amplitude of transit timing variations is known to be very large ($\sim$97 minutes), but with a $\sim$3400 year precession period.

CoRoT-7b is a very hot super-Earth and has the shortest known orbital period (excepting the ultra-short period planets
of \citealt{2006Natur.443..534S}). We included this planet in our analysis to get a feel for the plausibility of 
detecting the interior structure of terrestrial extra-solar planets. The small radius reduces the 
planetary contribution to apsidal precession (Figure 1) and significantly reduces the photometric signal. We note here
that in bodies where material strength (rigidity) is more important than self-gravity, $k_{2p}$ is no longer directly related to 
internal density distribution. The correction factor is typically small for bodies larger than the Earth \citep{1999ssd..book.....M}.

XO-3b is a super-massive eccentric planet that is not in the \emph{Kepler} field. Even so, it is interesting to note that, using the known
eccentricity $e=0.2884 \pm 0.0035$ \citep{2009arXiv0902.3461W} and accounting for the brightness of the host star (V=9.8), the \emph{Kepler} 
threshold $k_{2p}$ is reduced to only 0.54. As pointed out by \citet{JB08} and \citet{PK08}, XO-3b is a good candidate for observing 
apsidal precession within the next decade or so. Furthermore, as discussed below, the non-zero obliquity of the stellar spin axis \citep{2009arXiv0902.3461W} 
may also result in an observable signal due to nodal precession.

HAT-P-7b and HAT-P-11b are orbiting bright stars in the \emph{Kepler} field. The latter is an eccentric hot Neptune with a 
relatively large semi-major axis resulting in no eminently detectable apsidal precession. HAT-P-7b, on the other hand, is a good candidate
for detecting apsidal precession. It is probably one of the brightest 
hot Jupiters in the \emph{Kepler} field, orbiting a V=10.5 star. The system brightness improves the expected photometric accuracy 
from 1000 ppm/min to 200 ppm/min, implying that an eccentricity of only 0.014 is needed to detect apsidal precession (threshold $k_{2p}$=0.3). 
\citet{2009IAUS..253..428P} report a best-fit eccentricity of 0.003 $\pm$ 0.012, indicating that the necessary eccentricity cannot be ruled out. 
Furthermore, this planet has transiting data extending back to 2004 and was observed by NASA's \emph{EPOXI} Mission in 2008
(\citealp{2009IAUS..253..301C}; D. Deming, pers. comm.). This additional baseline, though sparsely sampled, may provide the additional
leverage needed to detect apsidal precession if the eccentricity is non-zero. 
Note, however, that detecting changes in transit shapes is more difficult when the observations
are made with a variety of telescopes because transit shapes depend on the observing filter used, due to wavelength-dependent 
limb darkening.

TrES-2b is similar to HAT-P-7b in that it also lies in the \emph{Kepler} field, has observations dating to 2005, and was observed by NASA's \emph{EPOXI} Mission. 
TrES-2b is somewhat fainter than HAT-P-7b (V=11.4), and, correcting for the system brightness, an eccentricity of 0.021 would result in detectable apsidal 
precession (threshold $k_{2p}$=0.3). Observations of the secondary eclipse show no detectable deviations of the orbit from circularity \citep{2009IAUS..253..536O}. Even so, 
the light curve of this planet is quite sensitive to perturbations as it has a quite high impact parameter $b=0.854$. Accounting for this impact parameter
does not significantly change the required eccentricity. 

We conclude that \emph{Kepler may detect the
  cores of very hot Jupiters} and probe their interior structure
though their evolving transit light curve if eccentricities are above $\sim$0.003. 
As future observations provide longer baselines
for these observations, the sensitivity to interior structure measurements
will increase dramatically, significantly lowering the eccentricity needed
to observe apsidal precession. 

In cases where apsidal precession is not
observed, the data can set strong upper limits on 
planetary eccentricities. An upper limit on the eccentricity
can be inferred by assuming that the planet
has the minimal physically-plausible value of $k_{2p} \approx 0.1$. 
Null detections of apsidal motion should therefore provide
upper limits on eccentricity comparable to the values shown in Table 1 (also shown by 
circles in Figure \ref{ek2planet}). Such strong eccentricity constraints are valuable 
for improving our understanding of these close-in planets. 

\section{Potential Confusion of the Apsidal Precession Signal}

In the above, we have assumed that measuring $\dot{\omega}$ is
tantamount to measuring $k_{2p}$. This is justified by noting that
the conversion $\dot{\omega}$ to $k_{2p}$ involves only factors that
are very well characterized. In Section 2 and Figure 1, we showed that
$k_{2p}$ is usually the dominant source of apsidal precession. 
The effects of $k_{2*}$ and general relativity are well-understood and 
can typically be subtracted away without introducing serious uncertainty, even when they dominate the apsidal precession
rate. From Equation \ref{tidalprec}, converting the remaining
$\dot{\omega}_{\rm p}$ to $k_{2p}$ requires only knowing $\frac{M_p}{M_*}$, $e$, $\frac{R_p}{a}$,
and $n$. The latter two are very accurately measured
with even a few transit light curves \citep[e.g.,][]{Torres2008, Southworth2008}.
The eccentricity only enters the equation through the $f_2(e)$ and $g_2(e)$
eccentricity functions (Equations \ref{f2e} and \ref{g2e}), and \emph{Kepler} observations of secondary eclipse
are sufficiently accurate to remove any systematic error due to these terms
unless the eccentricity is large ($e\gtrsim0.3$). 
Determining the mass ratio requires well-sampled radial 
velocity observations. The systems detected by \emph{Kepler} are
bright enough to get good mass measurements, especially
since very hot Jupiters have large radial velocity amplitudes ($K \sim 200$
m/s).\footnote{Other than determining the mass ratio and constraining the 
eccentricity, radial velocity
  information is thought to have a negligible contribution in constraining
  apsidal precession unless a serious observational campaign can 
measure the radial velocity period (independently of transits) to sub-second accuracies.
 \citep{2007MNRAS.377.1511H,JB08}.} 
The anticipated error in the mass
ratio is a few percent \citep{Torres2008}. In all, we estimate that,
converting from $\dot{\omega}$ to $k_{2p}$ leads to a typical systematic error on
$k_{2p}$ of around $\sim$.01. This is a relatively small systematic effect in comparison to the
potential range ($\sim$0.5) of $k_{2p}$ values. For reference, the eccentricity required to reach a 
threshold $k_{2p}$ of 0.01 is shown in Figure \ref{ek2planet} by squares.

Another way to introduce systematic errors on the measurement of $k_{2p}$ is 
to misinterpret similar transit light curve variations. To ensure that the method 
outlined in this paper truly probes the interiors of extra-solar 
planets, we consider in this section whether the transit light curve
resulting from apsidal 
precession can be confused with any other common circumstances. Although a 
very specific combination of parameters is required for any particular phenomenon
to successfully mimic a signal due to $k_{2p}$, the below effects 
should be reconsidered when actual data is available.

\subsection{Testing the Effect of Obliquity}
\label{obliquitysec}

If either the star or planet has a non-zero obliquity, the orbital plane will no longer
be fixed as a result of nodal precession. The obliquities of
very hot Jupiters rapidly ($\lesssim 1$ MYr) decay to a 
Cassini state, and recent work has shown that these planets are likely
in Cassini state 1 \citep{2005ApJ...628L.159W,2007A&A...462L...5L,2007ApJ...665..754F}.  
Using a model based on the equations of \citet{2001ApJ...562.1012E}, we found that 
Cassini obliquities of very hot Jupiters are indeed negligible ($\alpha_p < 0.01^{\circ}$).
Though tidal damping of the stellar obliquity occurs on far longer
timescales, 
several measurements of the projected stellar obliquity through the
Rossiter-McLaughlin effect indicates that planet-hosting stars generally have
low obliquities $\lesssim 10^{\circ}$ like the Sun \citep{2009arXiv0902.0737F}. 
Hence, the general expectation is that both the star and planet will have rather 
low, but potentially non-zero obliquities.

Understanding the specific orbital evolution resulting from non-zero
obliquities is more complicated than the simple prescription for
apsidal precession. 
To correctly account for non-Keplerian effects, we wrote a direct
integrator, following \citet{2002ApJ...573..829M}, that 
calculates the Cartesian trajectory (and the direction of the spin axes) of a star-planet 
system including general 
relativity and the effects of quadrupolar distortion. This integrator
reproduces the orbit-averaged analytic 
equations of \citet{2002ApJ...573..829M}, which are the same as those in
\citet{2001ApJ...562.1012E}, \citet{1939MNRAS..99..451S}, and
elsewhere.\footnote{This involved minor modifications to the "direct
  integrator" equations 3 and 5 in \citet{2002ApJ...573..829M}. In Equation 3, the coefficient 12 should 
be a 6 (R. Mardling, pers. comm.) and Equation 5 was replaced with the nearly equivalent
 equation from \citet{1989racm.book.....S}.} We did
not include the effects of tidal forces or additional planets which
are not relevant to our problem. 

Using this direct integrator, we investigated the effect of non-zero obliquities on the 
transit times, durations, and impact parameters. Integration of several cases with 
varying stellar and planetary obliquities showed that the largest effect on the 
photometry was due to changes in the impact parameter, as expected for an orbit with 
changing orientation \citep{2002ApJ...564.1019M}. However, even for large stellar 
obliquities ($\sim 45^{\circ}$) the transit light curve variations due to obliquity are 
generally small relative to the effects of purely apsidal precession, even with low 
eccentricities. One reason for this is that the tidal bulge, which does not contribute
to nodal precession, is $\gtrsim15$ times 
more important than the rotational bulge. 
As with apsidal precession, the planetary contribution to orbital variations 
is much stronger than the stellar contribution (for equal obliquities). Unless the 
planetary obliquity is unexpectedly large ($\gtrsim 0.5^{\circ}$), 
the obliquity-induced nodal precession should have only a 
minor effect on the transit light curve.

\subsection{Transit Timing due to Orbital Decay}
\label{tidessec}

Orbital decay generates a small secular trend in transit times. 
\citet{2003ApJ...596.1327S} proposed the detectability
of the expected $\sim$1 ms/yr period change due to semi-major axis decay 
of
OGLE-TR-56b. The transit timing anomaly due solely to orbital decay (or growth) is
the result of constantly accumulating changes in the period: 
\be\label{ttta}
T_N \simeq T_0 + NP_{\rm obs} + \frac{1}{2} N^2 \delta P
\ee
where $\delta P \equiv \dot{P}P$ is the change in the period during
one orbit and $N$ is the number of transits after the initial
transit. Equation \ref{ttta} can be derived by noting that the transit
times are 
basically the integral of the instantaneous period. As before, the 
transit timing anomaly is composed of the quadratic deviation of $T_N$ from a straight line. 
The change in period can be due to magnetic stellar breaking \citep[e.g.,][]{2009AJ....137.3181L,2009arXiv0902.4554B}, 
the Yarkovsky effect applied to planets \citep{2008ApJ...677L.117F}, and/or other effects. 

For planets orbiting an asynchronously rotating star, a major
source of orbital decay is tidal evolution, which results in a slow change in 
semi-major axis, according to the formula \citep{1999ssd..book.....M}:
\be
\dot{a} = \textrm{sign}(\nu_* - n) \frac{3k_{2*}}{Q_*} \frac{M_p}{M_*} \left( \frac{R_*}{a} \right)^5 n a
\ee
where sign($x$) returns the sign of $x$ or 0 if $x=0$ and where $Q_*$ is the tidal
 quality parameter of the star, typically around $\e{4}$ \citep{2004ApJ...610..464D}.
Though $\delta P$ due to 
tidal dissipation is only of order 3 micro-seconds, $N$ grows by $\sim 300$ each year,
reaching $\sim$1000 during the duration of \emph{Kepler} for
very hot Jupiters. This implies a transit timing signal of about a few seconds.

Calculating the total ``signal-to-noise'' of tidal evolution, as was
done for $k_{2p}$, we find that reasonable values of $Q_*$ can be
measured even 
for faint stars ($V=14$; 1000 ppm/min noise). For a circular 
orbit with the
parameters of OGLE-TR-56b, the effective $\frac{S}{N}$
reaches 1 when $\dot{P}$ is 1.36 ms/yr (see Table 1), corresponding to $Q_* \approx 25000$. This implies the 
detectability
of most of the empirically-motivated estimates of \citet{2003ApJ...596.1327S} for
the tidal decay of OGLE-TR-56b, which 
are estimated to be within an order of magnitude of 1 ms/yr. On the other hand, 
\citet{2009arXiv0902.4563B} estimate that the tidal damping in F-stars like OGLE-TR-56
and WASP-12 may be very low, which may explain the survival of these short period planets.

The estimates of the threshold values of $\dot{P}$, shown in Table 1, include
removing degeneracies in other parameters, except apsidal precession of
eccentric orbits, and assume that everything but $\dot{P}$ is known. Note that the
transit light curve signal due to orbital decay is due entirely to transit timing; the
change in $a$ is far too small to observe in transit shaping. As the signal
due to apsidal precession includes significant changes to the shapes of the
transits, the signal due to $k_{2p}$ is qualitatively different than that of $Q_*$.
The shifting of secondary transits from precession also help in this
regard, as outlined above. However,
the primary transit timing signals can be similar: quadratic transit timing
anomalies with amplitudes of $\sim$1 second. \label{threshpdot}

\emph{Kepler} analogs of very hot Jupiters WASP-12b, OGLE-TR-56b, CoRoT-1b, WASP-4b, and TrES-3b 
could have detectable transit timing anomalies due to tidal decay,
implying a direct measurement of the current value of $Q_*$ for specific stars (Table 1). This is an
exciting possibility, providing the first direct measurements (or constraints)
of the currently unknown details of tidal dissipation in
a variety of individual stars.\footnote{The vanishingly small effect of eccentricity decay 
is $\sim \frac{1}{Q_p}$ smaller than apsidal precession, so that
direct measurements of $Q_p$ from eccentricity decay are not feasible.} We also note that interesting orbital decay of
eclipsing binary systems seen by \emph{Kepler} could also be detectable.

\subsection{Confusion Due to Other Planets}

Could the signal due to $k_{2p}$ be confused with additional planets? In
considering this issue, it should be noted that all known hot Jupiters (with $a
\lesssim 0.05$ AU and $M_p \gtrsim 0.5 M_{Jup}$) have no currently known additional
companions. The apparent single nature of these systems could very
well be due to observational 
biases \citep{2008arXiv0806.4314F}. However, even for stars that
have been observed for many years with radial velocity (e.g. 51 Peg, HD
209458), there appears
to be a strong tendency towards hot Jupiters as the only close-in massive planets. 

Previous studies of transit timing variations focus on the effects of additional 
planetary perturbers \citep[e.g.,][]{2005Sci...307.1288H,2005MNRAS.359..567A,2007ApJ...664L..51F,2008ApJ...688..636N}. 
These authors find that nearby massive planets or even low-mass planets in mean-motion resonances
would cause strong transit timing variations that are easily distinguishable from the comparatively
long-period timing anomalies due to $k_{2p}$. Relatively distant
companions or non-resonant low-mass planets, however, can induce a linear apsidal precession signal 
just like $k_{2p}$ \citep{2002ApJ...564.1019M,2007MNRAS.377.1511H,JB08}. The precession rate
induced by a perturbing body is a function of its mass and semi-major axis. The interior structure of very hot Jupiters
causes apsidal precession as fast as a few degrees per year. To match this precession rate
would require, for example, another Jupiter-mass planet at $\lesssim 0.1$AU or a solar-mass star at $\sim$1 AU.
Even perturbers an order of magnitude smaller than these would be readily detectable using radial velocity observations
and/or high-frequency transit time variations. When restricted to planets that are undetectable by other means, 
adding the precession due to the unknown perturbing planet would lead to an 
insignificant overestimate of $k_{2p}$ for very hot Jupiters.\footnote{Conversely, as a consequence
of the fast precession of very hot Jupiters due to their (unknown) interiors, it 
will be very difficult to detect the presence of additional perturbing planets in these systems
from apsidal precession alone.} When observing transiting planets
with larger semi-major axes ($a \gtrsim 0.05$ AU), the strength of planetary
induced apsidal precession is reduced to a level comparable to apsidal precession
from a low-mass perturbing planet \citep{JB08} and confusion may be possible in these cases.

Since the transit timing signal for apsidal precession is similar to a
sinusoid, another potential source of confusion would be light-travel
time offsets due to a distant orbiting companion \citep[e.g.,][]{2008A&A...480..563D}. The transit timing signal due
to stellar motion about the barycenter can be distinguished from
$k_{2p}$ precession\footnote{Transit time anomalies due to $Q_*$ (Section
  \ref{tidessec}), however, can be confused with barycenter light-travel time
  shifts due to a distant planet that may be undetectable in radial velocities.} 
by considering the changes in transit shapes and
primary-secondary transit time offsets, which are not affected by distant companions. 

We conclude that transit timing effects from other planets can be readily
distinguished from the effects of apsidal precession. To address the issue of 
the transit shaping signal due to additional planets, we wrote a simple three-body
integrator (similar to the integrator mentioned above) to investigate the kinds of 
transit light curve signals created by additional planets. For the vast majority of additional planet parameters, 
the transit timing deviations always carry far more signal than the minor deviations
due to changes in the angular velocity\footnote{The angular velocity
is directly related to the star-planet separation through conservation
of angular momentum: $r\dot{f}^2$.} ($\dot{f}_{tr}$) or impact parameter ($b$), which together
determine the transit shape as described in Section \ref{transitshapes} above. Generally, 
it is much easier to delay
a transit by 5 seconds than it is to shift the apparent transit plane by an appreciable amount. 

However, when the perturbing planet is on a plane highly-inclined to the transiting planet, changes in the transit shape
can become detectable, even while the transit timing variations are negligible. For example, a perturbing 
planet of mass $\e{-5}M_*$ at 0.1 AU with a mutual inclination of 45$^{\circ}$ caused very hot Jupiter transit 
durations to change by $\sim$1 second/year. This kind of signal is the result of nodal precession induced
by the perturbing planet, as originally pointed out by \citet{2002ApJ...564.1019M}. In our investigation, 
we found that the three-body nodal precession alters the impact parameter ($b$) but does not significantly 
affect the orbital angular velocity ($\dot{f}_{tr}$). Conversely, the transit shaping signal due to $k_{2p}$ is 
generally produced by changes in both $b$ and $\dot{f}_{tr}$, but at near-central transits, the 
effect of changing orbital velocity is dominant (see Section \ref{transitshapes}). In high-precision 
transit light curves, both the angular velocity and the impact parameter can be 
independently measured and hence the signals of apsidal and nodal
precession are usually distinct for all but the most grazing transits.

Given the uniqueness of the apsidal precession signal induced by the planet's interior, it appears that if additional
planets are not detectable in radial velocities, transit timing variations, or nodal precession, then 
they will not contribute to a misinterpretation of an inferred value of $k_{2p}$ for very hot Jupiters. 
Nevertheless, future measurements of $k_{2p}$ should check that these issues are unimportant 
within the context of the specific system being studied.

Finally, we estimate that moons or rings with enough mass to bias an
inferred $k_{2p}$ would cause other readily detectable photometric
anomalies \citep[e.g., planet-moon barycentric motion][]{1999A&AS..134..553S}. 
In addition, extra-solar moons with any significant mass are tidally unstable,
especially around very hot Jupiters \citep{2002ApJ...575.1087B}.

\section{Other Methods for Determining $k_{2p}$}

\subsection{Secular Evolution of a Two Planet System}
Measuring $k_2$ for an extra-solar planet was suggested by \citet{2002ApJ...564.1024W}
for the inner planet of HD 83443. Unfortunately, later analyses have
indicated that the supposed second planet in this system was actually
an artefact of the sparse radial velocity data \citep{2004A&A...415..391M}. Nevertheless,
this technique could be applied to other eccentric planetary systems
with similar conditions \citep{2007MNRAS.382.1768M}. \citet{2002ApJ...564.1024W}
showed that in a 
regime of significant tidal circularization and 
excitation from an additional planet, the ratio of eccentricities
depends on the precession rate which is dominated by
$k_{2p}$ as shown above (see also \citealt{2006ApJ...649.1004A}, who do not
include precession due to the planetary quadrupole). In theory, the
current orbital state of such multi-planet systems gives an 
indirect measurement of the apsidal precession rate. 

\subsection{Direct Detection of Planetary Asphericity}
Another method for determining interior properties of transiting planets
would be to directly measure the asphericity due to the rotational or tidal
bulge in primary transit photometry. The height of the
rotational and tidal bulges are $q_r h_2 R_p$ and $q_t h_2 R_p$,
respectively, where $q_r$ and $q_t$ are the dimensionless small
parameters defined in Equation \ref{qparams} and $h_2$ is
another Love number which, for fluid bodies, is simply $k_2 + 1$ \citep{1939MNRAS..99..451S}. 
These bulges cause the disk of the planet to be slightly
elliptical, subtly modifying the photometric signal, as discussed for
rotational bulges by \citet{2002ApJ...574.1004S} and \citet{2003ApJ...588..545B}. However, as  
discussed by \citet{2003ApJ...588..545B}, in real systems with actual
observations, the size of the 
rotational bulge is very
difficult to determine as it is highly correlated with
stellar and orbital parameters that are not known 
\emph{a priori}, e.g. limb darkening coefficients.

The tidal bulge, whose height is also set by $k_{2p}$,
does not suffer from some of the difficulties involved
with measuring the rotational bulge. It has a known orientation
(pointing towards the star) so there is no degeneracy from an unknown
obliquity \citep{2003ApJ...588..545B}. (Note, however, that for 
hot Jupiters, the obliquities must be tidally evolved to nearly zero, so this 
isn't really a problem with the rotational bulge.) In addition, the signal due to oblateness is only
significant near ingress/egress, but the tidal bulge is continuously
changing orientation throughout the entire
transit. Though the tidal 
bulge is typically three times larger than the rotational bulge
(Equation 2), the projection 
of the tidal bulge that is seen during a transit is small, proportional to 
$\sin \theta$ where $\theta$ is the angle between the planet position
and the Earth's line of sight. For very hot Jupiters that have
semi-major axes of only $\lesssim$6 stellar radii, $\sin \theta$ during transit
ingress/egress 
reaches $\gtrsim \frac{1}{6}$ so that the projected tidal bulge is
about half as large as the rotational bulge. The extra dimming due to the tidal bulges (and rotational
bulges) is as high as $2 \x \e{-4}$ for some planets that are expected to
have tides over 
2000 km high (e.g. WASP-12b, WASP-4b,
Corot-1b, OGLE-TR-56b); this compares very favorably with
the photometric accuracy of binned \emph{Kepler} data at about
10 ppm per minute. However, we expect that, as with the 
rotational bulge alone, the combined signal from the rotational and tidal bulge 
will be highly
degenerate with the unknown limb-darkening coefficients, as the
size of the projection of the tidal bulge also varies as the distance to
the center of the star. 

We note that using multi-color photometry should significantly improve
the prospects of detecting non-spherical planetary transits since it
breaks most of these degeneracies. For
example, \citet{2007ApJ...655..564K} use HST to observe transits of HD 209458b in 10 wavelength
bands and measure the planetary radius with a relative accuracy
(between bands) of $0.003 R_J$, of the same level as the change in
shape due to oblateness and the tidal bulge. \citet{2007A&A...476.1347P} made a
similar measurement for HD 189733b and reached even higher relative
accuracy. Combining such
measurements with other data (e.g. primary transits in the infrared,
where limb-darkening is much smaller) and a stellar photosphere model (to
correctly correlate limb darkening parameters as in \citealt{2007MNRAS.374..941A})
 could yield detections 
of planetary asphericity, especially in very hot Jupiters which have
the largest bulges. 

One possible source of confusion in interpreting planetary asphericity is the thermally-induced pressure
effects of an unevenly radiated surface. In non-synchronous planets, the thermal tidal bulge \citep{2009arXiv0901.0735A} can 
shift the level of the photosphere by approximately an atmospheric scale height, about 
$\e{-2}$ or $\e{-3}$ planetary radii (P. Arras, pers. comm.). The orientation of the thermal bulge is significantly 
different from the tidal or rotational bulges and should be distinguishable. 
Furthermore, very hot Jupiters should orbit synchronously, reducing the importance of this effect. Nevertheless, the effect
of atmospheric phenomena on measurements of planetary asphericity should be considered. 

Though difficult to disentangle from other small 
photometric effects, high-precision multi-color photometry may be another viable
method for measuring $k_{2p}$. This technique is complimentary to detecting $k_{2p}$ 
from apsidal precession since it does not require that the planet is eccentric, 
nor does it require a long time baseline. On some planets, 
the two methods could be used together as mutual confirmation
 of the planetary interior structure.

\section{Conclusions}

The planetary mass and radius are the only bulk physical characteristics measured 
for extra-solar planets to date. In this paper, we find that the planetary Love number ($k_{2p}$, 
equivalent to $J_2$)
can also have an observationally detectable signal (quadrupole-induced apsidal precession) which can provide a new and 
unique probe into the interiors of very hot Jupiters. In particular, $k_{2p}$ 
is influenced by the size of a solid core and other internal properties. Core sizes can be used to infer the formation
and evolution of individual extra-solar planets \citep[e.g.,][]{2009arXiv0901.0582D,2009arXiv0903.1997H}.

The presence of a nearby massive star creates a large tidal potential on these 
planets, raising significant tidal bulges which then induce non-Keplerian effects on 
the star-planet orbit itself. The resulting apsidal precession accounts for $\sim$95\% of 
the total apsidal precession in the best cases (Figure \ref{odotfig}). Hence, we 
find that the internal density distribution, characterized by $k_{2p}$, has a large 
and clear signal, not to be confused with any other parameters or phenomena. We urge 
those modeling the interior structures of extra-solar planets to tabulate the
values of $k_{2p}$ for their various models.

Encouraged by this result, we calculated full photometric light-curves like those 
expected from the \emph{Kepler} mission to determine the realistic observability 
of the interior signal. We estimate that \emph{Kepler} should be able to 
distinguish between interiors with and without massive cores ($\Delta k_{2p} \simeq 
0.1$) for very hot Jupiters with eccentricities around $e \sim 0.003$ (Figure 
\ref{ek2planet}). Eccentricities this high may occur for some of the very 
hot Jupiters expected to be found by \emph{Kepler}, though
these planets usually have highly damped eccentricities. Much stronger constraints
on apsidal precession can be obtained by combining \emph{Kepler} photometry
with precise secondary transits observed in the infrared. In cases where apsidal precession is not
observed, the data can set strong upper limits on planetary eccentricities.

In analyzing \emph{Kepler}'s photometric signal of apsidal precession, we find that transit 
timing variations are an almost negligible source of signal, though transit timing 
has been the focus of many observational and theoretical papers to date. The effect 
of ``transit shaping'' has $\sim$30 times the photometric signal of transit timing for 
apsidal precession \citep[see Figure \ref{candywrapperpieces},][]{PK08,JB08}). At 
orientations where transit timing and shaping are weakest, the changing offset 
between primary and secondary transit times can be used to measure $k_{2p}$ (Figure 
\ref{bowtiemain}). It may also be possible to measure $k_{2p}$ from high-precision 
multi-color photometry by directly detecting the planetary asphericity in transit.
Such a measurement does not require a long baseline or an eccentric orbit.

Very hot Jupiters are also excellent 
candidates for detecting tidal semi-major axis decay, where we find that relatively 
small period changes of $\dot{P} \simeq 1$ ms/yr should be detectable. This could 
constitute the first measurements (or constraints) on tidal $Q_*$ for a variety of 
individual stars. We note that \emph{Kepler} measurements of transit timing and shaping 
for eclipsing binaries should also provide powerful constraints on stellar interiors through
apsidal motion and binary orbital decay (due to tides, if the components are asynchronous).

Accurately measuring the interior structure of distant extra-solar planets seems 
too good to be true. Nevertheless, the exquisite precision, constant monitoring, 
and 3.5-year baseline of the \emph{Kepler} mission combined with the high 
sensitivity of transit light curves to small changes in the star-planet orbit make 
this measurement plausible. 

Our focus on \emph{Kepler} data should not be interpreted to mean that other observations will 
be incapable of measuring $k_{2p}$. In fact, the opposite is true since 
the size of the apsidal precession signal increases 
dramatically with a longer baseline. Combining \emph{Kepler} measurements with future ground
and space based observations can create a powerful tool for measuring $k_{2p}$. 
In the far future, many planets will have 
measured apsidal precession rates (like eclipsing binary systems have now) and inferred 
$k_{2p}$ values. Incorporating these measurements into interior models holds
 promise for greater understanding of all extra-solar planets.

\begin{acknowledgements}
We thank Dave Stevenson, Mike Brown, Greg Laughlin, Oded Aharonson, Thomas Beatty, Phil Arras, 
Rosemary Mardling, Re'em Sari, Alejandro Soto, Ian McEwen 
and Chris Lee for help and useful discussions. We especially 
thank the referee, Dan Fabrycky, for helpful suggestions and discussions.
DR is grateful for the 
support of the Moore Foundation. ASW is grateful for support from the National Science 
Foundation. This research has made use of NASA's Astrophysics Data System.
\end{acknowledgements}

\end{document}